\documentclass{article}

\usepackage[english]{babel}
\usepackage{parskip}
\usepackage{bm}

\usepackage[letterpaper,top=2cm,bottom=2cm,left=3cm,right=3cm,marginparwidth=1.75cm]{geometry}

\usepackage{amsmath, amssymb, natbib, booktabs, amsfonts, amsthm}
\usepackage{dsfont}
\usepackage{graphicx}

\usepackage{amsmath,amsthm,amssymb,amsfonts,graphicx,dsfont}
\usepackage[maxfloats=100]{morefloats}
\usepackage[utf8]{inputenc}
\usepackage{booktabs}
\usepackage{epsfig}
\usepackage{url}
\usepackage{bm,setspace}
\usepackage{rotating}
\usepackage{listings}
\usepackage{verbatim}
\usepackage{xcolor}
\usepackage[title]{appendix}
\usepackage{multirow} 
\usepackage{longtable, graphicx}
\usepackage{adjustbox}
\usepackage{bbm}
\usepackage{float} 
\usepackage{subfigure, subcaption} 
\newtheorem{remark}{Remark}
\usepackage[colorlinks=true, allcolors=blue]{hyperref}

\newtheoremstyle{break}
  {\topsep}{\topsep}%
  {\itshape}{}%
  {\bfseries}{}%
  {\newline}{}%
\theoremstyle{break}

\newtheorem{proposition}{Proposition}
\newtheorem{exampleemph}[proposition]{Example}   % upshape

\usepackage{changepage}   % for the adjustwidth environment
\makeatletter
\newcommand*{\rom}[1]{\expandafter\@slowromancap\romannumeral #1@}
\makeatother

\title{Gated Graph Neural Networks for Learning Hidden Independent Cascade Dynamics}
\author{Anubha Goel, Illia Oleksiienko, Mateusz Wilinski, Alexandros Iosifidis, Juho Kanniainen}

\begin{document}
\maketitle
\begin{abstract}

Information and infectious diseases spread through social networks, but the spreading probabilities driving them are hard to estimate without per-node activation times. Applications seldom supply these, and inference must instead proceed through indirect and noisy proxies for the terminal infection states. We study this inverse problem on a fixed, known graph under the Hidden Independent Cascade (HIC) model, with one spreading probability per node rather than a single global rate, so the number of unknowns scales with the number of nodes. Seed sets and observation parameters are known, while activation times and terminal infection states are latent, and the observed-data likelihood requires marginalizing over every spreading outcome. We propose a simulation-based amortized estimator that recovers the full node-level parameter vector without reconstructing individual latent cascades. Repeated seed-conditioned symptom observations are summarized as Symptom-Aware Cascade Features (SACF), which combine empirical symptom statistics with neighborhood and structural information. SACF are mapped to parameters by SAGE-HC, a permutation-equivariant gated graph neural network whose learned gates attenuate neighborhood messages corrupted by false positives and false negatives, and training on simulated HIC realizations yields a reusable inverse map. As a benchmark under the same hidden observations, we extend the Dynamic Message Passing learning framework to the HIC emission model. On synthetic and empirical graphs the two methods separate by topology. DMP is highly accurate on trees, whereas SAGE-HC is substantially better on heterogeneous, loopy graphs under noisy terminal symptoms.

\end{abstract}

\section{Introduction}

Network models in general, and spreading dynamics on networks in particular, constitute a useful family of tools for modeling problems in epidemiology \citep{pastor2015epidemic}, finance \citep{acemoglu2015systemic}, information spreading \citep{zapperi2025epidemiology}, and many other fields \citep{newman2018networks, barrat2008dynamical}.
Many widely used spreading models have only a small number of node states, discrete-time evolution, and simple local stochastic rules.
One of the advantages of such models is their simplicity, which often permits direct computations and efficient Monte Carlo simulations.
Nevertheless, inverse tasks for spreading processes, like recovering the parameters of the dynamics from observations, can remain computationally demanding, especially when the spreading process is only indirectly observed.

In this work, we aim to estimate a separate spreading probability for every node from indirect observations made only after the cascade has terminated. The underlying dynamics follow the Independent Cascade model on a fixed and known directed graph, with one parameter per node shared across its outgoing edges, so the number of unknowns equals the number of nodes. The data record not the cascades themselves but what they emit, namely a known seed set and a noisy terminal symptom vector for each of a set of independent cascades. Neither the per-node activation times used by standard cascade-inference methods nor the terminal infection field is available. The parameters of the symptom-observation mechanism are assumed known, so the spreading parameters are the sole target of estimation.

These two sources of information loss are of different kinds. The first is temporal. Observations arrive only after the cascade has terminated, so there is no ordering to induce the increasing sequence of activated subgraphs that standard cascade data provide. The second concerns the state itself. Because the terminal infection field is seen only through noisy symptoms, a positively symptomatic node need not be connected to the seeds through a path of live edges, and an infected node may produce no symptom at all. The observations therefore contain both false positives and false negatives. Direct likelihood evaluation requires marginalizing over a large number of latent spreading outcomes, separately for every cascade, and the resulting objective must be optimized over as many unknowns as there are nodes.

In practice, complete cascade data can rarely be assumed. Our setting therefore arises naturally in applications where the underlying spreading state is latent and only noisy proxies are observed. Examples include
financial markets, where private information diffusion is inferred
from trading behavior or investor proximity
\citep{baltakys2019neighbors, baltakys2023predicting, goel2025topological};
epidemiology, where infections are latent and symptoms provide
imperfect signals \citep{braunstein2016inference, allen2020population};
cybersecurity, where alerts act as proxies for compromised systems
\citep{munoz2017efficient, kazeminajafabadi2023optimal}; and
online information diffusion, where latent adoption is observed
through engagement or resharing behavior
\citep{bakshy2009social, taylor2013selection}.

Early work on inverse problems for spreading processes focused on fully observable cascades and showed how maximum-likelihood methods can be used to learn model parameters or reconstruct the underlying network \citep{saito2008prediction, gomez2012inferring, pouget2015inferring}.
In many applications, however, activation times are not recorded at all, and learning must rely on the final outcome of each cascade \citep{amin2014learning}, where guarantees for recovering the network hold for tree-like graphs but weaken as short cycles become common.
The observations may also be incomplete, with some nodes unobserved throughout, and they may themselves be noisy.
Direct maximum-likelihood inference then becomes computationally expensive because it requires summation over the unobserved part of the system \citep{lokhov2016reconstructing}.
One way to approximate this computation is to use Dynamic Message Passing (DMP) as a subroutine for learning \citep{wilinski2021prediction}.
The resulting Scalable Learning of Independent Cascade Effective Representation framework was shown to handle several forms of noise and incompleteness in cascade data \citep{wilinski2024learning}, and provides the natural likelihood-based comparison for learning spreading parameters from noisy and partially observed data.
More generally, message passing is a widely used approximation for both forward and inverse inference in network dynamics \citep{karrer2010message, altarelli2014bayesian, li2021impact, crotti2023matrix}, including reconstruction tasks relevant to our work \citep{braunstein2019network}.

Several other methods have addressed inference from incomplete cascade observations.
In \citep{dutta2018bayesian}, likelihood-free approximate Bayesian computation (ABC) is used to jointly infer a small set of spreading parameters and the source from multiple fully observed snapshots.
Extending this line of work, \cite{wang2024flexible} used neural-network-compressed ABC to learn informative summary statistics and infer spreading parameters on a known network from a sequence of positive and negative tests, without observing infection and removal times.
An autoregressive neural-network method within a Bayesian framework was proposed in \citep{biazzo2022bayesian}, but its computational cost restricts its applicability to relatively small networks.
Independent Cascade network inference from probabilistic terminal infection statuses was studied in \citep{huang2024learning}, but focused on reconstructing unknown influence relationships rather than estimating node-level spreading parameters on a known graph.
More recently, \cite{manoharan2025learning} introduced the Hidden Independent Cascade (HIC) model and proposed a likelihood-free procedure in which a classifier is trained to distinguish observed cascades from simulated ones, while the model parameters are adjusted until the two become indistinguishable.
Although effective for low-dimensional parameterizations, this approach does not scale effectively to the heterogeneous setting considered here, in which the number of unknown parameters grows with the number of nodes.

We retain the HIC generative mechanism introduced in \citep{manoharan2025learning}, but instantiate its spreading layer with a heterogeneous node parameterization. In contrast to the low-dimensional likelihood-free estimation considered there, our goal is to infer a separate spreading probability for every node from repeated observations of known seeds and noisy terminal symptoms. We address this problem using a graph-based estimator trained on simulated HIC realizations, which shares information across nodes through the graph rather than treating each parameter on its own.

Our main contributions are as follows. First, we provide a formal mathematical
formulation of the heterogeneous HIC inverse problem under noisy terminal
observations, including the latent spreading process, observation mechanism,
and observed-data likelihood. Second, we introduce \emph{Symptom-Aware Cascade
Features (SACF)}, which aggregate repeated hidden-cascade observations across
cascades and seed configurations into structured node-level summaries. SACF
combine symptom frequencies and variability with neighborhood co-activation,
graph-distance, and self-seeded spreading information, without reconstructing
individual latent cascades. Third, we propose the \emph{Symptom-Aware Gated
Estimator for Hidden Cascades (SAGE-HC)}, a permutation-equivariant graph
neural network based on gated message passing.
Training SAGE-HC on simulated spreading-parameter assignments and corresponding
HIC observations yields a form of simulation-based amortized inference
\citep{zammit2025neural}, learning a reusable inverse map from graph-aware
cascade summaries to node-level spreading parameters. Fourth, we extend the
Dynamic Message Passing learning framework of
\citep{wilinski2021prediction, wilinski2024learning} to the HIC observation
model, providing a likelihood-based benchmark under the same hidden terminal
observations. Finally, we systematically examine the effects of network
topology, observation noise, training data, cascade budget, repeated feature
realizations, and test-time observation budgets on estimation accuracy and
stability.

\section{Model and Method}

Our approach consists of three stages. We first specify the HIC spreading
and observation model, then construct SACF from the observed cascades, and
finally use SAGE-HC to estimate the node-level spreading parameters.

% Our approach consists of three stages.
% First, we formalize spreading using the Hidden Independent Cascade (HIC)
% model, which specifies a latent stochastic spreading process and a noisy
% observation mechanism.
% Second, we construct Symptom-Aware Cascade Features (SACF), which map the
% observed symptom data into node-level statistics aligned with the structure
% of the HIC model.
% Finally, conditioning on these features, we learn node susceptibilities
% using a graph-based regression model called Symptom-Aware Gated Estimator for Hidden Cascades (SAGE-HC) that exploits network structure.
% This separation between modeling, feature construction, and learning
% allows us to leverage probabilistic insight while retaining flexibility
% in the estimator.

\subsection{Hidden Independent Cascade: Problem Formulation}

The Independent Cascade (IC) model is a standard probabilistic model of
spreading on networks, in which activated nodes attempt to activate each
neighbor once, independently, according to fixed transmission probabilities.
In many applications, however, activations are not directly observed; instead,
only noisy terminal signals (e.g., symptoms, test results, or behavioral
traces) are available as imperfect proxies for infection. The HIC model
extends IC with an explicit observation layer: the spreading process follows
IC dynamics, while the terminal infection state remains latent and only noisy
emissions are observed.

Let $(\Omega,\mathcal F,\mathbb P_\Theta)$ be a probability space supporting
the spreading and observation processes, where $\Theta$ indexes the induced
law $\mathbb P_\Theta$ and comprises the diffusion and observation parameters.
Let $G=(V,E)$ be a finite directed graph with $n=|V|<\infty$ nodes and
$E\subseteq V\times V$ directed edges. We observe $M$ independent cascades,
indexed by $m=1,\ldots,M$, with known deterministic seed sets
$S^{(m)}\subseteq V$ activated at $t=0$.

In the general formulation, each directed edge $(i,j)\in E$ has a
transmission probability $\theta_{ij}\in[0,1]$. A parsimonious special case
is obtained by setting
$
\theta_{ij}=p_i,
$
where $p_i\in[0,1]$ is shared across all outgoing edges of node $i$.
At each time $t\geq1$, every node activated at $t-1$ attempts to activate
each out-neighbor exactly once, with transmission along $(i,j)$ succeeding
with probability $\theta_{ij}$, independently across edges and attempts.
The dynamics are progressive, so nodes are activated at most once; since
$G$ is finite and each edge is used at most once, every cascade terminates
in finite time.

For each cascade $m$ and edge $(i,j)\in E$, define the edge-success
random variable
\[
X_{ij}^{(m)}:\Omega\to\{0,1\},
\qquad
X_{ij}^{(m)}\sim\mathrm{Bernoulli}(\theta_{ij}),
\]
with independence across edges and cascades. Here $X_{ij}^{(m)}=1$
indicates that edge $(i,j)$ is live in cascade $m$, while
$X_{ij}^{(m)}=0$ indicates that it is blocked. For each cascade $m$, define the terminal infection field
\[
\mathbf Z^{(m)}
=(Z_i^{(m)})_{i\in V}:\Omega\to\{0,1\}^V,
\]
where $Z_i^{(m)}=1$ if node $i$ becomes infected during the cascade and
$Z_i^{(m)}=0$ otherwise. In the live-edge representation,
$Z_j^{(m)}=1$ whenever $j$ is reachable from $S^{(m)}$ through a directed
path of live edges. In particular,
$
Z_i^{(m)}=1,\forall i\in S^{(m)}.
$
We refer to $\mathbf Z^{(m)}$ as the latent infection field and denote its
generated $\sigma$-algebra by
$\sigma(\mathbf Z^{(m)})\subseteq\mathcal F$.
Its state space is $\{0,1\}^V$, equipped with the product $\sigma$-algebra,
which coincides with the power set since $V$ is finite.

% Edge-success random variables: for each cascade $m$ and each edge
% $(i,j)\in E$, we define
% \[
% X_{ij}^{(m)}:\Omega\to\{0,1\},
% \qquad
% X_{ij}^{(m)}\sim\mathrm{Bernoulli}(\theta_{ij}),
% \]
% mutually independent across $(i,j)\in E$ and independent across cascades
% $m$. We interpret $X_{ij}^{(m)}=1$ as edge $(i,j)$ being live (able to
% transmit in cascade $m$), and $X_{ij}^{(m)}=0$ as blocked. For each cascade $m$, we define \[
% \mathbf{Z}^{(m)}=(Z_i^{(m)})_{i\in V}:\Omega\to\{0,1\}^V,
% \]
% where $Z_i^{(m)}=1$ indicates that node $i$ becomes infected at some time
% during cascade $m$, and $Z_i^{(m)}=0$ otherwise. More precisely $X_{ij}^{(m)}=1$ leads to infection state $Z_j^{(m)}=1$. We call $\mathbf{Z}^{(m)}$ a latent infection field. Observe that, by definition of seeds,
% \[
% Z_i^{(m)}=1 \quad \text{for all } i\in S^{(m)}. 
% \]
% %\color{red}{\mathbb{P}_\Theta\text{-a.s.}}
% We denote the $\sigma$-algebra generated by $\mathbf{Z}^{(m)}$ by
% $\sigma(\mathbf{Z}^{(m)})\subseteq\mathcal{F}$. The terminal latent infection configuration for a cascade is represented by
% a binary vector in $\{0,1\}^V$, equipped with its product $\sigma$-algebra
% (which coincides with the power set since $\{0,1\}^V$ is finite). 
We observe the symptom vector
\[
\mathbf{s}^{(m)}=(s_i^{(m)})_{i\in V}:\Omega\to\mathcal{X}^V,
\]
where $s_i^{(m)}$ is a noisy proxy for the terminal infection status of node
$i$. Here $(\mathcal{X},\mathcal{B}(\mathcal{X}))$ is a measurable
observation space (e.g., $\mathcal{X}=\{-1,0,+1\}$ or $\mathcal{X}=\{0,1\}$),
and $\mathcal{X}^V$ is equipped with the product $\sigma$-algebra
$\mathcal{B}(\mathcal{X})^{\otimes V}$.

The infection field $\mathbf{Z}^{(m)}$ is latent, and
$\mathbf{s}^{(m)}$ is generated from it through a noisy observation model.
Conditional on the deterministic seed sets $\{S^{(m)}\}_{m=1}^M$, the pairs
$\{(\mathbf{Z}^{(m)},\mathbf{s}^{(m)})\}_{m=1}^M$ are independent under
$\mathbb{P}_\Theta$. Conditional on $\mathbf{Z}^{(m)}$, symptoms are
independent across nodes and depend only on the corresponding infection state.
Thus, for every $A\in\mathcal{B}(\mathcal{X})$,
\[
\mathbb{P}_{\Theta_{\mathrm{obs}}}
\!\left(s_i^{(m)}\in A\mid Z_i^{(m)}=z_i\right)
=\nu_{z_i}(A),
\]
and, for any finite $U\subseteq V$ and measurable $A_i\subseteq\mathcal{X}$,
\[
\mathbb{P}_{\Theta_{\mathrm{obs}}}\!\left(
\bigcap_{i\in U}\{s_i^{(m)}\in A_i\}
\,\middle|\,\mathbf{Z}^{(m)}
\right)
=
\prod_{i\in U}
\mathbb{P}_{\Theta_{\mathrm{obs}}}
\!\left(s_i^{(m)}\in A_i\mid Z_i^{(m)}\right).
\]

% Next, we define 
% \[
% \mathbf{s}^{(m)}=(s_i^{(m)})_{i\in V}:\Omega\to\mathcal{X}^V,
% \]
% where $s_i^{(m)}$ is an imperfect proxy for the terminal infection status
% of node $i$ in cascade $m$. The
% observed symptom configuration lies in $\mathcal{X}^V$, where
% $(\mathcal{X},\mathcal{B}(\mathcal{X}))$ is a measurable observation space
% (e.g., $\mathcal{X}=\{-1,0,+1\}$ or $\mathcal{X}=\{0,1\}$), and $\mathcal{X}^V$
% carries the product $\sigma$-algebra $\mathcal{B}(\mathcal{X})^{\otimes V}$. 

% We assume that the infection field $\mathbf{Z}^{(m)}$ is not directly observed. Instead, we observe a symptom vector $\mathbf{s}^{(m)}$ generated as a noisy
% emission from $\mathbf{Z}^{(m)}$.
% Finally, conditional on the deterministic seed sets $\{S^{(m)}\}_{m=1}^M$, we assume that the pairs
% $\{(\mathbf{Z}^{(m)},\mathbf{s}^{(m)})\}_{m=1}^M$ are independent under $\mathbb{P}_\Theta$, i.e., cascades are i.i.d.\ replicates of the same HIC process. Given this conditional independence, symptoms are independent
% across nodes and each node's symptom distribution depends only on its own
% infection status. That is, there exist probability measures such that for all measurable
% $A\in\mathcal{B}(\mathcal{X})$,
% \[
% \mathbb{P}_{\Theta_{\mathrm{obs}}}(s_i^{(m)}\in A \mid Z_i^{(m)}=z_i)=\nu_{z_i}(A),
% \]
% and for any finite $U\subseteq V$ and measurable sets $A_i\subseteq\mathcal{X}$,
% \[
% \mathbb{P}_{\Theta_{\mathrm{obs}}}\!\left(\bigcap_{i\in U}\{s_i^{(m)}\in A_i\}\ \middle|\ \mathbf{Z}^{(m)}\right)
% =
% \prod_{i\in U}\mathbb{P}_{\Theta_{\mathrm{obs}}}(s_i^{(m)}\in A_i\mid Z_i^{(m)}).
% \]

\begin{remark}\textbf{Tri-valued symptom model:}
In the tri-valued case $\mathcal{X}=\{-1,0,+1\}$, infected nodes emit a
positive symptom with probability $q\in(0,1]$ and are otherwise neutral:
\[
\mathbb{P}(s_i^{(m)}=+1 \mid Z_i^{(m)}=1)=q,\quad
\mathbb{P}(s_i^{(m)}=0 \mid Z_i^{(m)}=1)=1-q,\quad
\mathbb{P}(s_i^{(m)}=-1 \mid Z_i^{(m)}=1)=0.
\]
Uninfected nodes generate false/noisy symptoms according to a categorical
distribution $(\eta_+,\eta_0,\eta_-)$:
\[
\mathbb{P}(s_i^{(m)}=+1 \mid Z_i^{(m)}=0)=\eta_+,\quad
\mathbb{P}(s_i^{(m)}=0 \mid Z_i^{(m)}=0)=\eta_0,\quad
\mathbb{P}(s_i^{(m)}=-1 \mid Z_i^{(m)}=0)=\eta_-,
\]
where $\eta_+ + \eta_0 + \eta_- = 1$. In this example $\Theta_{\mathrm{obs}}=(\eta_+, \eta_0, \eta_-,q)  $
\end{remark}
\begin{remark}
We assume $\nu_0\neq\nu_1$, ensuring that symptoms contain information
about latent infection. If $\nu_0=\nu_1$, then $\mathbf{s}^{(m)}$ is
independent of $\mathbf{Z}^{(m)}$ and spreading parameters are not
identifiable from symptom-only data.
\end{remark}

Given the observed cascades
$\mathcal D=\{(S^{(m)},\mathbf s^{(m)})\}_{m=1}^M$, where
$S^{(m)}\subseteq V$ is the seed set and
$\mathbf s^{(m)}\in\mathcal X^V$ denotes the node-level observations,
the observed-data likelihood, assuming independence across cascades, is
\[
L(\Theta;\mathcal D)
=\prod_{m=1}^M
\mathbb P_\Theta(\mathbf s^{(m)}\mid S^{(m)}),
\qquad
\Theta=(\Theta_{\mathrm{diff}},\Theta_{\mathrm{obs}}).
\]
Direct evaluation of
$\mathbb P_\Theta(\mathbf s^{(m)}\mid S^{(m)})$ is computationally
intractable because it requires summation over all possible spreading
outcomes. Since the latent infection state $\mathbf Z^{(m)}$ is unobserved,
it is marginalized from the joint distribution:
\[
\mathbb P_\Theta(\mathbf s^{(m)}\mid S^{(m)})
=
\sum_{z\in\{0,1\}^V}
\mathbb P_\Theta\!\left(
\mathbf Z^{(m)}=z,\mathbf s^{(m)}=s
\mid S^{(m)}
\right).
\] 

Under the assumed two-stage generative process, $\mathbf Z^{(m)}$ is first
drawn from the spreading model
$\mathbb P_{\Theta_{\mathrm{diff}}}(\cdot\mid S^{(m)})$ and, conditional on
$\mathbf Z^{(m)}=z$, symptoms are generated independently across nodes as
$\mathbb P_{\Theta_{\mathrm{obs}}}(\mathbf s^{(m)}=s\mid
\mathbf Z^{(m)}=z)=\prod_{i\in V}\nu_{z_i}(\{s_i\})$.
Hence,
\begin{align}
\mathbb P_\Theta(\mathbf Z^{(m)}=z,\mathbf s^{(m)}=s\mid S^{(m)})
&=
\mathbb P_{\Theta_{\mathrm{diff}}}(\mathbf Z^{(m)}=z\mid S^{(m)})
\prod_{i\in V}\nu_{z_i}(\{s_i\}).
\end{align}
Substituting into the marginal likelihood gives
\[
L(\Theta;\mathcal D)
=
\prod_{m=1}^M
\sum_{z\in\{0,1\}^V}
\mathbb P_{\Theta_{\mathrm{diff}}}(\mathbf Z^{(m)}=z\mid S^{(m)})
\prod_{i\in V}\nu_{z_i}\big(\{s_i^{(m)}\}\big).
\]

Direct likelihood-based inference remains challenging because the terminal
infection states are latent and must be marginalized over stochastic
diffusion outcomes, while the observed symptoms introduce additional
measurement noise. We therefore estimate the node-level spreading parameters
using a graph-based regression model trained on simulated HIC realizations.

\subsection{Feature Engineering: Symptom-Aware Cascade Features}

\textbf{Why not use the raw symptom observations?} In the Hidden Independent Cascade model, the latent infection states
$\mathbf Z^{(m)}$ are unobserved, and learning relies on terminal symptom
vectors $\mathbf s^{(m)}$. Raw per-cascade observations are highly variable:
each cascade is a stochastic realization of the spreading process, while
$s_u^{(m)}$ is a noisy proxy for $Z_u^{(m)}$. Thus, raw symptoms combine
spreading variability with observation noise. We therefore aggregate
observations across multiple cascades and seed configurations into
\emph{Symptom-Aware Cascade Features (SACF)}, yielding more stable node-level
summaries that are informative about the underlying spreading parameters.
% In the Hidden Independent Cascade model, the latent infection states
% $\mathbf Z^{(m)}$ are unobserved, and learning must rely on terminal
% symptom vectors $\mathbf s^{(m)}$. Treating raw, per-cascade symptom values
% as learning inputs is statistically inappropriate for two reasons. First, each cascade constitutes a single random realization of the
% underlying spreading process. Even conditional on fixed transmission
% parameters, the event $Z_u^{(m)}=1$ is stochastic, so single-cascade
% observations provide high-variance samples of node-level behavior. Second, symptoms are generated through a noisy observation channel: $s_u^{(m)}$ is only an imperfect proxy for $Z_u^{(m)}$, with nonzero
% false-positive and false-negative rates. Using raw symptoms therefore
% entangles spreading variability with measurement noise. To address both sources of variability, we aggregate symptom observations
% across multiple cascades and seed configurations and construct
% \emph{Symptom-Aware Cascade Features (SACF)}, which summarize the observed
% data into stable node-level statistics that are informative about the
% latent spreading process while remaining robust to observation noise and
% finite-sample variability. 

%\subsubsection{Feature-map formalization}

Given the observed dataset
$\mathcal D=\{(S^{(m)},\mathbf s^{(m)})\}_{m=1}^{M}$ and graph $G$, we
associate each node $u\in V$ with a feature vector
$\mathbf x_u=\phi_u(\mathcal D;G)\in\mathbb R^d$.
Each coordinate of $\phi_u$ is constructed from the observed symptoms,
seed sets, and network structure. These features are not intended to
reconstruct the latent states $Z_u^{(m)}$ of individual cascades, but to
summarize observed spreading patterns that are informative about the
node spreading parameter $p_u$. In the following, we construct $\phi_u$ for
the tri-valued symptom model.

% \subsubsection{Feature-map formalization}

% Let
% \[
% \mathcal D \in (S \times \mathcal X^V)^M
% \]
% denote the observed dataset, viewed as a random element under
% $\mathbb P_\Theta$.
% We define a node-level feature map
% \[
% \phi : (S \times \mathcal X^V)^M \times G \;\longrightarrow\; \mathbb R^d,
% \qquad
% u \longmapsto \phi_u(\mathcal D;G),
% \]
% where each coordinate of $\phi_u$ is a measurable function of the observed
% symptoms, seed sets, and network structure.

% The role of $\phi_u$ is not to recover the latent infection indicators
% $Z_u^{(m)}$, but rather to approximate low-order statistics of the
% observed-data likelihood that are informative about the node susceptibility
% parameter $p_u$.

% When desired, we similarly define pairwise features as a measurable map
% \[
% \psi: (S \times \mathcal X^V)^M \times G \to \mathbb R^{d_e},
% \qquad
% \psi_{uv}(\mathcal D;G)\in \mathbb R^{d_e},\ \ (u,v)\in V\times V,
% \]
% which can be restricted to $(u,v)\in E$ if only directed edges are used.
% In our implementation, we primarily use node features $\phi_u$; the
% definition of $\psi_{uv}$ clarifies how symptom-derived correlations can
% be formalized when edge features are included. We model the symptoms using the tri-valued model.

\paragraph{Distance-weighted symptom statistics.}
Fix a collection of seeds $\{s_k\}_{k=1}^K$ and let
$\mathcal C_k := \{m:S^{(m)}=\{s_k\}\}$ denote the cascades initiated from
seed $s_k$. To account for attenuation with graph distance, define
\[
w(u,s_k)=\exp\!\big(-\alpha d(u,s_k)\big),
\]
where $d(u,s_k)$ is the shortest-path distance from $s_k$ to $u$, and
$w(u,s_k)=0$ if $u$ is unreachable. The corresponding weighted symptom
counts are
\[
\begin{aligned}
N_{u,+}^{(k)} &:= \sum_{m\in\mathcal C_k}
\mathbf 1\{s_u^{(m)}=+1\}\,w(u,s_k), &
N_{u,-}^{(k)} &:= \sum_{m\in\mathcal C_k}
\mathbf 1\{s_u^{(m)}=-1\}\,w(u,s_k),\\
N_{u,0}^{(k)} &:= \sum_{m\in\mathcal C_k}
\mathbf 1\{s_u^{(m)}=0\}\,w(u,s_k).
\end{aligned}
\]
These quantities provide distance-adjusted empirical symptom frequencies
at node $u$.

\paragraph{Neighborhood symptom intensity.}
To capture local correlations induced by spreading, define
\[
\Lambda_u^{(m)}
:=\frac{1}{d_u}\sum_{v\in N(u)}
\mathbf 1\{s_v^{(m)}=+1\},
\qquad d_u:=|N(u)|,
\]
which measures the fraction of neighbors of $u$ exhibiting positive symptoms
in cascade $m$. To capture local co-activation, define the conditional statistics
\[
\Lambda_{u\mid +}^{(m)}
:=
\begin{cases}
\dfrac{1}{d_u}\displaystyle\sum_{v\in N(u)}
\mathbf 1\{s_v^{(m)}=+1\},
& d_u>0,\ s_u^{(m)}=+1,\\
0, & \text{otherwise.}
\end{cases}
\]
For seed $s_k$, the corresponding unconditional and conditional empirical
means are
\[
\bar\Lambda_u^{(k)}
:=\frac{1}{|\mathcal C_k|}
\sum_{m\in\mathcal C_k}\Lambda_u^{(m)},
\qquad
\bar\Lambda_{u\mid +}^{(k)}
:=
\frac{\sum_{m\in\mathcal C_k}\Lambda_{u\mid +}^{(m)}}
{\max\!\left(1,\sum_{m\in\mathcal C_k}
\mathbf 1\{s_u^{(m)}=+1\}\right)}.
\]

\paragraph{Cross-seed aggregation.}
Since different seeds excite different regions of the network, we aggregate
the seed-specific features as
\[
N_{u,+}:=\frac{1}{K}\sum_{k=1}^{K}N_{u,+}^{(k)},\quad
N_{u,-}:=\frac{1}{K}\sum_{k=1}^{K}N_{u,-}^{(k)},\quad
N_{u,0}:=\frac{1}{K}\sum_{k=1}^{K}N_{u,0}^{(k)},
\]
\[
\bar{\Lambda}_u:=\frac{1}{K}\sum_{k=1}^{K}\bar{\Lambda}_u^{(k)},
\qquad
\bar{\Lambda}_{u\mid +}:=
\frac{1}{K}\sum_{k=1}^{K}\bar{\Lambda}_{u\mid +}^{(k)}.
\]

\paragraph{Symptom uncertainty via entropy.}
Let $\widehat{\pi}_u$ denote the empirical symptom distribution at node $u$
computed from $(N_{u,+},N_{u,-},N_{u,0})$. Its entropy is
\[
H_u:=-\sum_{x\in\mathcal X}\widehat{\pi}_u(x)\log \widehat{\pi}_u(x).
\]

% \begin{remark}[Uncertainty interpretation]
% High entropy indicates unstable or weakly excited nodes for which observed
% symptoms provide limited information about the latent infection state,
% explicitly exposing epistemic uncertainty induced by sparse or noisy data.
% \end{remark}
\paragraph{Self-seed spreading statistics.}
Cross-seed aggregation may obscure the spreading pattern generated when node
$u$ itself acts as the source. For a seeded node $u=s_k$ and
$m\in\mathcal C_k$, define
\[
A_u^{(m)}
:=
\begin{cases}
\dfrac{1}{d_u}\displaystyle\sum_{v\in N(u)}
\mathbf 1\{s_v^{(m)}=+1\}, & d_u>0,\\
0, & d_u=0.
\end{cases}
\]
Here, $A_u^{(m)}$ is the fraction of immediate neighbors exhibiting a positive
symptom when the cascade is initiated from $u$. We summarize its average level
and variability across self-seeded cascades as
\[
\mu^{\mathrm{nbr}}_{u,\mathrm{self}}
:=\frac{1}{|\mathcal C_k|}\sum_{m\in\mathcal C_k}A_u^{(m)},
\qquad
\left(\sigma^{\mathrm{nbr}}_{u,\mathrm{self}}\right)^2
:=\frac{1}{|\mathcal C_k|}
\sum_{m\in\mathcal C_k}
\left(A_u^{(m)}-\mu^{\mathrm{nbr}}_{u,\mathrm{self}}\right)^2.
\]

To capture propagation beyond the immediate neighborhood, define
\[
R_u^{(m)}
:=\frac{1}{|V|-1}
\sum_{\substack{v\in V\\v\neq u}}
\mathbf 1\{s_v^{(m)}=+1\},
\]
the fraction of non-seed nodes exhibiting a positive symptom in cascade $m$.
Its mean and variance across self-seeded cascades are
\[
\mu^{\mathrm{reach}}_{u,\mathrm{self}}
:=\frac{1}{|\mathcal C_k|}\sum_{m\in\mathcal C_k}R_u^{(m)},
\qquad
\left(\sigma^{\mathrm{reach}}_{u,\mathrm{self}}\right)^2
:=\frac{1}{|\mathcal C_k|}
\sum_{m\in\mathcal C_k}
\left(R_u^{(m)}-\mu^{\mathrm{reach}}_{u,\mathrm{self}}\right)^2.
\]

These four self-seed features characterize the local and global spreading
patterns generated by node $u$: the means capture the typical observed
response, while the variances quantify its variability across repeated
cascades. They therefore complement the cross-seed statistics by retaining
seed-specific information lost through aggregation.

When self-seeded cascades are unavailable, zero-valued self-seed statistics
must be distinguished from an observed response that is genuinely zero. We
therefore define
$I_u^{\mathrm{seed}}
:=\mathbf 1\{u\in\{s_k\}_{k=1}^K\}$, 
and set the four self-seed statistics to zero whenever
$I_u^{\mathrm{seed}}=0$.

The resulting Symptom-Aware Cascade Feature vector is
\[
\mathbf x_u=\phi_u(\mathcal D;G)=
\left(
\begin{aligned}
&N_{u,+},N_{u,-},N_{u,0},H_u,
\bar\Lambda_u,\bar\Lambda_{u\mid+},d_u,
d_{\min}(u,S),\overline d(u,S),\\
&\mu^{\mathrm{nbr}}_{u,\mathrm{self}},
(\sigma^{\mathrm{nbr}}_{u,\mathrm{self}})^2,
\mu^{\mathrm{reach}}_{u,\mathrm{self}},
(\sigma^{\mathrm{reach}}_{u,\mathrm{self}})^2,
I_u^{\mathrm{seed}}
\end{aligned}
\right)
\in\mathbb R^{14}.
\]
where $d_{\min}(u,S)$ and $\overline d(u,S)$ are the minimum nonzero and
average distances, respectively, from $u$ to the selected seed nodes.

\begin{remark}[Approximate sufficiency]
SACF are not sufficient statistics for $\Theta$; rather, they provide
low-dimensional summaries that reduce realization-level variability while
retaining information useful for spreading parameter estimation.
\end{remark}

For each spreading parameter assignment $\mathbf p$, we generate repeated
independent SACF realizations by resimulating HIC cascades, where each
realization consists of node features $\{\mathbf x_u\}_{u\in V}$ paired with
the same target spreading parameter vector $\mathbf p$. Full synthetic
data-generation details are provided in Appendix~\ref{app:data_generation}. Each SACF realization is a stochastic summary computed from finitely many
cascades. Using a very large cascade set for a single realization would reduce
its sampling variability, but would also produce a highly averaged
representation and reduce the diversity of training observations. We instead
generate multiple independent realizations per assignment using a moderate
number of cascades, providing repeated stochastic views of the same target
parameters while retaining both cascade-level variation and useful spreading
information. These repeated realizations also allow us to assess prediction stability.
For a fixed assignment, the trained model is evaluated on all independently
generated SACF realizations, and the resulting predictions are summarized by
the mean and standard deviation for each $p_i$. The corresponding intervals
provide an empirical measure of stability: narrow intervals indicate
consistent estimates across cascade realizations, whereas wider intervals
indicate greater sensitivity to limited or noisy cascade information.

\textbf{Learning from SACF.}
The SACF representation $\{\phi_u(\mathcal D;G)\}_{u\in V}$ provides
node-level summaries of the observed spreading outcomes while preserving
the relational structure of the underlying graph.
The remaining task is to learn a mapping from these features and the graph
topology to node susceptibilities.
Since the information used to estimate $p_u$ is contained in both the
node-level summaries and the surrounding neighborhood patterns, we employ a
graph-based learning architecture to capture these relational dependencies.

\subsection{Symptom-Aware Gated Estimator for Hidden Cascades (SAGE-HC)}

Given SACF node features $\{\mathbf x_u\}_{u\in V}$ and graph $G$, we learn
a graph-based estimator
\[
f_\theta:\big(G,\{\mathbf x_u\}_{u\in V}\big)
\longmapsto \{\hat p_u\}_{u\in V},
\]
where $\hat p_u$ estimates the node spreading parameter $p_u$. Since node labels
are arbitrary, $f_\theta$ is permutation-equivariant: relabeling the graph
and its node features induces the same relabeling of the predictions.

%\subsubsection{Residual gated message passing}

We use the residual gated graph convolution of
\citep{bresson2017residual}. With
$\mathbf h_u^{(0)}=\mathbf x_u$, layer $\ell$ updates node $u$ as
\begin{equation}
\mathbf h_u^{(\ell+1)}
=
\mathbf W_{\mathrm{self}}^{(\ell)}\mathbf h_u^{(\ell)}
+
\sum_{v\in\mathcal N(u)}
\boldsymbol{\eta}_{uv}^{(\ell)}
\odot
\mathbf W_{\mathrm{nbr}}^{(\ell)}\mathbf h_v^{(\ell)},
\label{eq:rggcn_final2}
\end{equation}
where
\begin{equation}
\boldsymbol{\eta}_{uv}^{(\ell)}
=
\operatorname{sigmoid}\!\left(
\mathbf W_u^{(\ell)}\mathbf h_u^{(\ell)}
+
\mathbf W_v^{(\ell)}\mathbf h_v^{(\ell)}
\right)
\in(0,1)^{d_{\ell+1}} .
\label{eq:gate_final2}
\end{equation}
Here, $\odot$ denotes element-wise multiplication. The residual term preserves
node-local SACF information, while the learned gates adaptively modulate
neighbor messages according to the representations of $(u,v)$. This is
particularly suited to HIC observations, where neighborhood symptom signals
can be informative but corrupted by false-positive and false-negative
observations; gating allows such contributions to be attenuated rather than
aggregated uniformly.

\begin{remark}[Why gating is natural for hidden cascades]
In HIC, neighborhood signals extracted from symptoms can be informative but
corrupted by false-positive and false-negative observations. The multiplicative
gates $\boldsymbol{\eta}_{uv}^{(\ell)}$ therefore provide an adaptive mechanism
for modulating neighbor contributions according to the current representations
of $(u,v)$, allowing the model to emphasize informative interaction pathways
rather than aggregate all neighbor messages uniformly.
\end{remark}

We construct a depth-$L$ stack of residual gated layers, each followed by
normalization and a pointwise nonlinearity. Let
$\mathrm{ResGatedConv}_{d_\ell\to d_{\ell+1}}$ denote the transformation in
\eqref{eq:rggcn_final2}--\eqref{eq:gate_final2}. For
$\ell=0,\dots,L-1$,
\begin{equation}
\mathbf h_u^{(\ell+1)}
=
\varphi\!\left(
\mathrm{Norm}\!\left(
\mathrm{ResGatedConv}_{d_\ell\to d_{\ell+1}}
(\mathbf h^{(\ell)})_u
\right)\right),
\label{eq:stack_update_final2}
\end{equation}
where $\varphi$ is ReLU and $\mathrm{Norm}$ denotes batch normalization.\footnote{Batch normalization stabilizes feature scales across layers for SACF
computed from cascade realizations with varying signal-to-noise ratios and
improves optimization on heterogeneous graphs.} The final embedding $\mathbf h_u^{(L)}\in\mathbb R^{d_L}$ is mapped to a
scalar spreading parameter estimate through a linear head,
$\tilde p_u=\mathbf w^\top \mathbf h_u^{(L)}+b.$
The unconstrained output is used during training and clipped to $[0,1]$ for
evaluation. Further architecture, optimization, normalization, and
data-splitting details are provided in
Appendix~\ref{app:implementation}. We refer to the resulting model as
\textbf{SAGE-HC} (Symptom-Aware Gated Estimator for Hidden Cascades).

% \textbf{Note:} For supervised learning, we generate synthetic training instances by
% sampling susceptibilities $\{p_u\}_{u\in V}$, simulating $M$ cascades under
% the HIC model, and constructing SACF inputs
% $\{\mathbf x_u=\phi_u(\mathcal D;G)\}_{u\in V}$ from the resulting symptom
% snapshots. Each instance thus provides a labeled pair
% $\big(\{\mathbf x_u\}_{u\in V},\{p_u\}_{u\in V}\big)$ on the fixed graph $G$.
% Training data consist of many independently generated such instances, enabling end-to-end learning of $f_\theta$. In our experiments, the parameters governing the observation model,
% $\Theta_{\mathrm{obs}}$, are assumed to be known. The learning task is
% therefore restricted to inference of the diffusion parameters
% $\Theta_{\mathrm{diff}}$ from symptom-level observations.

% \begin{figure}[t]
% \centering
% % \includegraphics[width=0.9\linewidth]{sage-hc-architecture.pdf}
% \caption{
% \textbf{SAGE-HC architecture.}
% Symptom-Aware Cascade Features (SACF) are first projected through a
% node-wise multilayer perceptron and aggregated locally.
% The resulting embeddings are propagated through multiple gated message
% passing layers, which selectively weight neighbor contributions and
% enable stable multi-hop information flow.
% A final linear head maps node embeddings to susceptibility estimates
% $\hat p_u$.
% }
% \label{fig:sage-hc}
% \end{figure}
\subsection{Dynamic Message Passing}

We adapt a Dynamic Message Passing (DMP) framework for learning spreading
parameters from partially observed Independent Cascade dynamics to the HIC
setting, where only terminal noisy symptoms are observed. The resulting
estimator is used as a likelihood-based baseline in our experiments.
Details are provided in Appendix~\ref{eq:dmp}.

\section{Results: Benchmarking with DMP}
%\subsection{Benchmarking with DMP}

We compare SAGE-HC with the HIC-adapted DMP baseline described above.
Experiments are conducted on five graph topologies: a balanced tree, three
Barabasi-Albert graphs with attachment parameters $m=2,3,4$, denoted
BA(2), BA(3), and BA(4), and the Karate Club network. These graphs span
different structural regimes, from an acyclic topology to increasingly
heterogeneous and loopy networks.

For each graph, we sample node-level spreading parameter vectors
$\mathbf{p}=(p_u)_{u\in V}$ and simulate HIC cascades to construct the
corresponding 14-dimensional SACF representation. The benchmark uses
$5{,}000$ spreading parameter assignments, with $10$ independent feature
realizations per assignment and $50$ cascades per seed. Every node is used
as a seed. We consider two observation regimes: \emph{noise-free}, with
$q=1$ and $\eta_{+}=\eta_{-}=0$, and \emph{hidden/noisy}, with
$q=0.6$ and $\eta_{+}=\eta_{-}=0.2$. SAGE-HC is trained on pairs $(\mathbf{X},\mathbf{p})$ and evaluated on
spreading parameter assignments entirely unseen during training. The same
held-out instances are used to evaluate DMP. Details of the assignment-aware
split, feature normalization, and checkpoint selection are provided in
Appendix~\ref{app:implementation}. We report Smooth L1 loss with $\beta=0.1$ and Acc@0.1, defined as the
fraction of predictions satisfying
$|\hat{p}_u-p_u^\star|<0.1$. SAGE-HC results are reported as
mean $\pm$ standard deviation over five independent training runs, whereas
DMP is deterministic for a fixed test dataset and is reported without
run-to-run variability.

\begin{table*}[ht!]
\centering
\small
\caption{Performance for the benchmark setting under
\textbf{noise-free} and \textbf{noisy} observations. Each SAGE-HC entry reports
\textbf{Smooth L1 / Acc@0.1}. Values are mean $\pm$ standard deviation across 5 
independent training runs.}
\label{tab:benchmark_performance_5000}

\renewcommand{\arraystretch}{1.10}
\setlength{\tabcolsep}{4pt}

\begin{tabular}{l|c|c}
\toprule
\textbf{Graph} & \textbf{DMP} & \textbf{SAGE-HC} \\
\midrule

\multicolumn{3}{c}{\textbf{(A) Noise-free observations}} \\
\midrule

Balanced tree
& $0.01154$ / $0.93012$
& $0.00872 \pm 0.00001$ / $0.95874 \pm 0.00016$ \\

BA(2)
& $0.18616$ / $0.33141$
& $0.00708 \pm 0.00004$ / $0.97814 \pm 0.00033$ \\

BA(3)
& $0.25926$ / $0.24653$
& $0.01086 \pm 0.00008$ / $0.95082 \pm 0.00081$ \\

BA(4)
& $0.28200$ / $0.20722$
& $0.01672 \pm 0.00008$ / $0.90181 \pm 0.00059$ \\

Karate Club
& $0.21747$ / $0.28993$
& $0.01404 \pm 0.00013$ / $0.92962 \pm 0.00123$ \\

\midrule
\multicolumn{3}{c}{\textbf{(B) Noisy observations}} \\
\midrule

Balanced tree
& $0.03138$ / $0.79754$
& $0.04722 \pm 0.00002$ / $0.67938 \pm 0.00075$ \\

BA(2)
& $0.19508$ / $0.32034$
& $0.01496 \pm 0.00005$ / $0.91884 \pm 0.00029$ \\

BA(3)
& $0.26045$ / $0.24241$
& $0.01732 \pm 0.00005$ / $0.89817 \pm 0.00056$ \\

BA(4)
& $0.28311$ / $0.20544$
& $0.02703 \pm 0.00004$ / $0.82264 \pm 0.00150$ \\

Karate Club
& $0.23057$ / $0.26292$
& $0.02852 \pm 0.00007$ / $0.80971 \pm 0.00082$ \\

\bottomrule
\end{tabular}
\end{table*}

The results indicate a non-monotonic relationship between graph structure and SAGE-HC performance. The balanced tree performs well in the noise-free setting but deteriorates substantially under observation noise, suggesting that its limited structural redundancy makes message passing more sensitive to corrupted terminal evidence. In loopy graphs, multiple paths and overlapping neighborhoods provide additional supporting information, which can help compensate for local observation noise. However, this benefit does not increase indefinitely: among the BA graphs, performance gradually degrades as connectivity increases from BA(2) to BA(4). Denser neighborhoods introduce more competing transmission paths and a larger amount of potentially noisy information to aggregate, making it harder to isolate the contribution of individual nodes.

DMP shows the complementary pattern. Its approximation is particularly well suited to the acyclic structure of the tree, but becomes less accurate as the graph contains more alternative and dependent transmission paths. SAGE-HC benefits from some degree of structural redundancy, but excessive connectivity can also make node-level inference more ambiguous. Thus, the strongest SAGE-HC performance appears at an intermediate level of connectivity, where the graph provides sufficient redundant evidence without introducing excessive mixing of noisy neighborhood information. Since increasing the BA parameter changes both density and loop structure, we interpret this trend as an effect of overall connectivity rather than attributing it solely to loopiness.

\paragraph{Prediction Stability and Sensitivity Analysis.}

In addition to the benchmark comparison, we examine the stability of SAGE-HC
with respect to the stochasticity of the observed cascades. For each unseen
test assignment, the underlying spreading parameter vector is fixed while the model
is evaluated on repeated feature realizations generated from independent
cascade samples. Variation in the resulting predictions therefore reflects
sensitivity to the particular cascade observations available at test time.
Across all graph structures, prediction variability is higher under noisy
observations, with the strongest increase observed for the balanced tree.
This indicates that corrupted terminal symptoms make spreading parameter estimates
more dependent on the specific cascade realization used for inference.
Detailed stability measures and results are reported in
Appendix~\ref{app:uncertainty}.

We further study the sensitivity of SAGE-HC to the amount of training and
observation data by varying the number of spreading parameter assignments,
cascades per seed, graph size, and feature realizations used during training.
Increasing the number of assignments and cascades produces the largest
improvements in the low- to moderate-data regimes, after which performance
shows clear diminishing returns. Repeated feature realizations also improve
estimation, particularly under noisy observations, although the gains become
small once the stochastic variability of an assignment is sufficiently
represented. Across the graph sizes considered, estimation accuracy does not
deteriorate as the number of node-level parameters increases.

Finally, we examine whether a model trained under an observation-rich setting
can generalize to substantially smaller cascade budgets at inference without
retraining. Although performance decreases when only a few cascades are
available, the model retains meaningful predictive accuracy even at strongly
reduced test-time budgets. Performance improves rapidly as additional cascade
evidence becomes available, with much of the gain achieved by moderate
cascade counts and diminishing improvements thereafter. The effect of limited
test-time information is stronger under noisy observations, although the gap
between the two observation regimes narrows as the cascade budget increases.
Complete results are provided in Appendix~\ref{app:sensitivity}.

\section{Conclusion}

We studied heterogeneous parameter inference for spreading processes when
only the seeds and noisy terminal symptoms are observed, while the infection
states and transmission paths remain latent. We first formulated the HIC model
mathematically for this hidden-observation setting, including the latent
spreading process, noisy observation mechanism, and observed-data likelihood.
Building on this formulation, we introduced SACF, which summarizes repeated
hidden-cascade observations into graph-aware node features, and SAGE-HC, a
gated graph neural network that learns an amortized mapping from these
observations to node-level spreading parameters without reconstructing
individual latent cascades. We also adapted Dynamic Message Passing to the HIC
observation model, providing a likelihood-based benchmark for the same
inference problem.

The experiments reveal a clear dependence on network structure and observation
quality. DMP is particularly effective on the tree, whereas SAGE-HC performs
better on heterogeneous, loopy networks, especially under noisy observations.
The stability and sensitivity analyses further show that observation noise
increases prediction variability, while additional parameter assignments,
cascade evidence, and repeated feature realizations improve estimation with
diminishing returns. A model trained with an observation-rich cascade budget
also remains effective at substantially smaller test-time budgets.

The present framework assumes a fixed known network and known observation
parameters. Joint inference of spreading and observation parameters, transfer
across related network structures, and applications where the latent spreading
parameters are unavailable remain natural directions for future work.

\subsection*{Funding} This work is supported by European Union's Horizon Europe programme under the Marie Skłodowska-Curie Actions (Grant Agreement No. 101150609, Project: HiddenTipChains).

\bibliographystyle{abbrv}
\bibliography{main}
\newpage
\section{Appendix-I}
\label{eq:dmp}

\subsection{General case with equal outgoing probabilities}

The objective function does not differ from the original work \citep{wilinski2021prediction, wilinski2024learning}.
\begin{equation}
\mathcal{O} = \sum_{s \in S}\sum_{i \in V}\sum_{\tau_i^s} m^{\tau_i^s}\,\log\!\big(p^s_i(\tau_i^s) \cdot \mathds{1}_{(\tau_i^s \le T)} - p^s_i(\tau_i^s - 1) \cdot \mathds{1}_{(\tau_i^s > 0)} + \mathds{1}_{(\tau_i^s = \ast)}\big),
\end{equation}
where $m^{\tau_i^s}$ counts the number of times node $i$ was activated at time $\tau_i^s$ for a cascade, which originated from $s$.
Function $p^s_i(t)$ represents the marginal probability of node $i$ being activated at time $t$ or before, for a cascade originating from $s$.
Note that $\tau_i^s = \ast$ means that the activation happened beyond our time horizon $T$, which may mean $\tau_i^s = \infty$ (node was never activated).
In other words, in this case we do not know the activation time, but we know that it did not happen in the time horizon.
Moreover, we are going to use a convention that $(\ast - 1) = T$.
As a result, for $\tau_i^s = \ast$ we get the probability of such an outcome equal to $1 - p^s_i(T)$, which is in line with our definition.
The constraints are slightly changed from the original formulation, due to the spreading probabilities being dependent only on the source node $\alpha_{ij} = \alpha_i$:
\begin{equation}
\begin{split}
\mathcal{C} &= \sum_{s \in S}\sum_{t=0}^{T}\sum_{j \in V} 
\lambda_j^s(t+1)\left[
p_j^s(t+1) - 1 + (1 - \bar p_j^s)\prod_{k \in \partial_{\mathrm{in}} j}
\left(1 - \alpha_k\,p_{k \to j}^s(t)\right)
\right] \\
&\quad + \sum_{s \in S}\sum_{t=0}^{T}\sum_{(i,j)\in E}
\lambda_{i \to j}^s(t+1)\left[
p_{i \to j}^s(t+1) - 1 + (1 - \bar p_i^s)\prod_{k \in \partial_{\mathrm{in}} i \setminus \{j\}}
\left(1 - \alpha_k\,p_{k \to i}^s(t)\right)
\right],
\end{split}
\end{equation}
where $\lambda_j^s(t)$ and $\lambda_{i \to j}^s(t)$ are the Lagrange multipliers.
This leads to the following Lagrange function:
\begin{equation}
    \mathcal{L} = \mathcal{O} + \mathcal{C}.
\end{equation}
Now we compute the derivatives, which will allow us to find the multipliers and compute the gradient over the probabilities.
\begin{equation}
\begin{split}
\frac{\partial \mathcal{L}}{\partial p_j^s(t)} 
&= \lambda_j^s(t)
+ \sum_{\tau_j^s} m^{\tau_j^s}\,
\frac{\mathds{1}_{(t=\tau_j^s)}\,\mathds{1}_{(\tau_j^s \le T)}}
     {p_j^s(\tau_j^s) - p_j^s(\tau_j^s-1)\,\mathds{1}_{(\tau_j^s > 0)}} \\
&\quad + \sum_{\tau_j^s} m^{\tau_j^s}\,
\frac{\mathds{1}_{(t=\tau_j^s-1)}\,\mathds{1}_{(\tau_j^s > 0)}}
     {p_j^s(\tau_j^s-1) - p_j^s(\tau_j^s)\,\mathds{1}_{(\tau_j^s \le T)} - \mathds{1}_{(\tau_j^s = \ast)}}.
\end{split}
\end{equation}
Equating left hand side to zero, allows to compute $\lambda_j^s(t)$.
\begin{equation}
\begin{split}
\frac{\partial \mathcal{L}}{\partial p_{a \to b}^s(t)}
&= \lambda_{a \to b}^s(t) \\
&\quad - \lambda_b^s(t+1)\,(1 - \bar p_b^s)\,
\alpha_a \prod_{m \in \partial_{\mathrm{in}} b \setminus \{a\}}
\left(1 - \alpha_m\,p_{m \to b}^s(t)\right) \\
&\quad - \sum_{k \in \partial_{\mathrm{out}} b \setminus \{a\}}
\lambda_{b \to k}^s(t+1)\,(1 - \bar p_b^s)\,
\alpha_a \prod_{m \in \partial_{\mathrm{in}} b \setminus \{a,k\}}
\left(1 - \alpha_m\,p_{m \to b}^s(t)\right).
\end{split}
\end{equation}
Equating left hand side to zero and using already known $\lambda_j^s(t)$, allows to compute $\lambda_{b \to k}^s(t)$.
Finally we can compute the derivative over the probabilities $\alpha_i$, which is equal to the gradient.
\begin{equation}
\begin{split}
\frac{\partial \mathcal{L}}{\partial \alpha_i}
&= -\sum_{s \in S}\sum_{t=0}^{T}\sum_{j \in \partial_{\mathrm{out}} i}
\lambda_j^s(t+1)\,(1 - \bar p_j^s)\,
\left[\, p_{i \to j}^s(t)\prod_{m \in \partial_{\mathrm{in}} j \setminus \{i\}}
\left(1 - \alpha_m\,p_{m \to j}^s(t)\right)\right] \\
&\quad -\sum_{s \in S}\sum_{t=0}^{T}\sum_{j \in \partial_{\mathrm{out}} i}
\sum_{k \in \partial_{\mathrm{out}} j \setminus \{i\}}
\lambda_{j \to k}^s(t+1)\,(1 - \bar p_j^s)\,
\left[\, p_{i \to j}^s(t)\prod_{m \in \partial_{\mathrm{in}} j \setminus \{i,k\}}
\left(1 - \alpha_m\,p_{m \to j}^s(t)\right)\right].
\end{split}
\end{equation}
The latter can be further simplified to:
\begin{equation}
\frac{\partial \mathcal{L}}{\partial \alpha_i}
= -\frac{1}{\alpha_i}\sum_{s \in S}\sum_{t=0}^{T-1}\sum_{j \in \partial_{\mathrm{out}} i}
\lambda_{i \to j}^s(t)\;p_{i \to j}^s(t),
\end{equation}
but only under the condition that $(\alpha_i \neq 0)$.

\subsection{Version for initial and final state data only}

Now we change the scenario, so that we only observe the initial and final state:
\begin{equation}
\mathcal{O} = \sum_{s\in S}\sum_{i\in V}
\Big[  m^{s}_{i,1}\,\log\!\big(p_i^s(T)\big)
\;+\; m^{s}_{i,0}\,\log\!\big(1 - p_i^s(T)\big)\; \Big],
\end{equation}
where $m^{s}_{i,0}$ and $m^{s}_{i,1}$ represent the number of times node $i$ was active or inactive, respectively, for a cascade with initial condition $s$.
This leads to a following change in the partial derivative over $p_j^s(t)$.
\begin{equation}
\frac{\partial \mathcal{L}}{\partial p_j^s(t)}
\;=\; \lambda_j^s(t)
\;+\; \mathds{1}_{(t=T)}\,\Bigg(\frac{m^{s}_{j,1}}{p_j^s(T)} \;-\; \frac{m^{s}_{j,0}}{1-p_j^s(T)}\Bigg).
\end{equation}
The other equations remain unchanged and are only affected by $\lambda_j^s(t)$ having different values then before.

\subsection{Hidden cascade version}

Final scenario assumes that we know the initial state and the symptoms of the final state.
Possible symptoms are positive (+), negative (-), and no symptoms (0).
For activated node, the probability of showing symptoms is $q$, while the probability of not showing symptoms is $1-q$.
For inactive node, the probability of showing symptoms is $a_1$, the probability of showing negative symptoms is $a_2$, and the probability of not showing symptoms is $1-a_1-a_2$.
As a result, the probabilities $\pi_{i,+}^s$, $\pi_{i,-}^s$, and $\pi_{i,0}^s$, of showing symptoms, negative symptoms, or no symptoms, respectively, for a cascade with initial condition $s$, are equal to:
\begin{equation}
\begin{split}
\pi_{i,+}^s &= q_i\,p_i^s(T) \;+\; a_{1,i}\,\big(1-p_i^s(T)\big)
             \;=\; a_{1,i} + (q_i-a_{1,i})\,p_i^s(T), \\
\pi_{i,-}^s &= 0 \cdot p_i^s(T) + a_{2,i}\,\big(1-p_i^s(T)\big) = a_{2,i}\,\big(1-p_i^s(T)\big), \\
\pi_{i,0}^s &= (1-q_i)\,p_i^s(T) \;+\; \big(1-a_{1,i}-a_{2,i}\big)\,\big(1-p_i^s(T)\big).
\end{split}
\end{equation}
The new objective is then:
\begin{equation}
\mathcal{O} \;=\; \sum_{s\in S}\sum_{i \in V}
\Big\{
m^{s}_{i,+}\,\log\!\big(\pi_{i,+}^s\big)
\;+\; m^{s}_{i,-}\,\log\!\big(\pi_{i,-}^s\big)
\;+\; m^{s}_{i,0}\,\log\!\big(\pi_{i,0}^s\big)
\Big\},
\end{equation}
where $m^{s}_{i,+}$, $m^{s}_{i,-}$, and $m^{s}_{i,0}$ represent the number of times node $i$ exhibited symptoms, negative symptoms, or no symptoms, respectively, for a cascade with initial condition $s$.
This leads to a following change in the partial derivative over $p_j^s(t)$.
\begin{equation}
\frac{\partial \mathcal{L}}{\partial p_j^s(t)}
\;=\; \lambda_j^s(t)
\;+\; \mathds{1}_{(t=T)}\;\Delta_j^s,
\end{equation}
where:
\begin{equation}
\begin{split}
\Delta_i^s
&= m^{s}_{i,+}\,\frac{\partial \pi_{i,+}^s / \partial p_i^s(T)}{\pi_{i,+}^s}
  \;+\; m^{s}_{i,-}\,\frac{\partial \pi_{i,-}^s / \partial p_i^s(T)}{\pi_{i,-}^s}
  \;+\; m^{s}_{i,0}\,\frac{\partial \pi_{i,0}^s / \partial p_i^s(T)}{\pi_{i,0}^s} \\[6pt]
&= m^{s}_{i,+}\,\frac{q_i-a_{1,i}}{\pi_{i,+}^s}
   \;-\; m^{s}_{i,-}\,\frac{a_{2,i}}{\pi_{i,-}^s}
   \;-\; m^{s}_{i,0}\,\frac{q_i-a_{1,i}-a_{2,i}}{\pi_{i,0}^s}.
\end{split}
\end{equation}
The other equations remain unchanged, like in the previous scenario.

\section{Appendix-II:}
% This appendix provides the architecture, optimization, normalization, and
% data-splitting details used in all SAGE-HC experiments.

\subsection{Synthetic data generation}
\label{app:data_generation}
Training data are generated on a fixed graph \(G=(V,E)\). Each assignment
consists of a sampled spreading parameter vector
\(\mathbf p=(p_u)_{u\in V}\), which serves as the ground-truth spreading
parameter. Conditional on \(\mathbf p\), we simulate \(M\) independent HIC
cascades \citep{hansen_baltakys_2026_cascadesimulator} and obtain symptom observations under the observation model
\(\Theta_{\mathrm{obs}}\). The resulting observations \(\mathcal D\) are
converted into SACF node features
\[
\mathbf x_u=\phi_u(\mathcal D;G), \qquad u\in V,
\]
forming a supervised sample \((\mathbf X,\mathbf p)\), where
\(\mathbf X=\{\mathbf x_u\}_{u\in V}\).

For a fixed assignment \(\mathbf p\), we may generate several feature
realizations
\[
\mathbf X^{(r)} \sim P(\mathbf X\mid \mathbf p,G,\Theta_{\mathrm{obs}}),
\qquad r=1,\ldots,R,
\]
by resimulating independent sets of \(M\) cascades. These realizations share
the same target \(\mathbf p\) but differ due to the stochasticity of the cascade
and observation processes. This is important because SACF features computed
from finitely many cascades are noisy empirical summaries of the latent
spreading process. Including multiple realizations for the same assignment
therefore trains the model to learn spreading parameter estimates that are stable
with respect to finite-sample cascade variability, rather than estimates that
depend on a single random cascade realization.

Repeating this procedure over many independently sampled assignments gives the
supervised training set. The observation parameters \(\Theta_{\mathrm{obs}}\)
are fixed and known in our experiments; hence, the learning task is to infer
\(\Theta_{\mathrm{diff}}\), specifically, the node spreading parameter
\(\mathbf p\), from symptom-level observations.

\subsection{Implementation Details}
\label{app:implementation}

\paragraph{Architecture.}
We use $d_0=d$ and $(d_1,d_2,d_3)=(256,128,64)$ as the default width
schedule (so $L=3$), although the formulation allows arbitrary depth and
widths.

\paragraph{Training protocol.}
Model parameters are optimized using Adam \citep{kingma2014adam} with a
learning rate of \(10^{-3}\), weight decay of \(10^{-4}\), and a batch size
of \(8\). The training objective is the Smooth L1 loss with parameter \(\beta=0.1\) 
\[
\ell(x,y)=
\begin{cases}
\frac{1}{2}\frac{(x-y)^2}{\beta}, & \text{if } |x-y|<\beta,\\[6pt]
|x-y|-\frac{1}{2}\beta, & \text{otherwise.}
\end{cases}
\]
Gradients are
stabilized by clipping their \(\ell_2\)-norm to \(1.0\) before each optimizer
update. All models are trained for \(100\) epochs, and the checkpoint with
the lowest validation selection loss is retained for evaluation. Input features are standardized independently by feature using
z-score normalization. For each feature \(j\), the mean \(\mu_j\) and
standard deviation \(\sigma_j\) are estimated exclusively from
the training samples, and
\[
\widetilde{x}_{ij}=\frac{x_{ij}-\mu_j}{\sigma_j}
\]
is applied unchanged to all training, validation, and test samples. Hence,
the held-out validation and test assignments do not contribute to the
normalization statistics.

To ensure evaluation on genuinely unseen spreading parameter configurations, we
use an assignment-aware hybrid split. Specifically, \(80\%\) of the
spreading parameter assignments are allocated to the development set, while
\(10\%\) and \(10\%\) are held out as unseen-validation and unseen-test
assignments, respectively. Within each development assignment, \(10\%\) of
the feature realizations are reserved for seen validation and the remaining
realizations are used for training; no development realizations are used for
testing. Checkpoint selection is based on the weighted validation criterion
\[
\mathcal{L}_{\mathrm{val}}
=
0.8\,\mathcal{L}_{\mathrm{val,seen}}
+
0.2\,\mathcal{L}_{\mathrm{val,unseen}},
\]
thereby accounting for both within-assignment generalization and
generalization to previously unseen spreading parameter assignments. Final
performance is reported exclusively on the held-out unseen-test assignments.
Each configuration is evaluated over five independent training runs.

\subsection{Empirical uncertainty estimation}
\label{app:uncertainty}

In addition to point-estimation performance, we examine the stability of
SAGE-HC with respect to the stochasticity of the observed cascades. This is
particularly relevant in the hidden-cascade setting, since the true node
susceptibilities are unavailable in a real application and the accuracy of an
estimated spreading parameter vector cannot be assessed directly. Repeated feature
realizations provide a simple way to quantify how sensitive the inferred
susceptibilities are to the finite cascade sample used to construct the input
features.

Recall that each assignment corresponds to a fixed latent spreading parameter
vector \(\mathbf p\), whereas its feature realizations are generated from
independent sets of cascades under the same \(\mathbf p\). Thus, variation in
the predictions across realizations reflects variation induced by the observed
cascade sample rather than a change in the underlying spreading parameter

For a fixed unseen-test assignment \(a\), let
\(\hat p_{a,i,j}\) denote the predicted spreading parameter of node \(i\) from
feature realization \(j=1,\ldots,n_a\). We first compute the mean prediction
\[
\bar p_{a,i}
=
\frac{1}{n_a}
\sum_{j=1}^{n_a}\hat p_{a,i,j}
\]
and the corresponding sample standard deviation \(s_{a,i}\) across
realizations. Following the standard normal-reference construction, we use
\[
\bar p_{a,i}
\pm
1.96\frac{s_{a,i}}{\sqrt{n_a}}
\]
as an empirical prediction-stability interval, with the endpoints restricted
to the admissible spreading parameter range \([0,1]\). The resulting interval width
for node \(i\) under assignment \(a\) is denoted by \(w_{a,i}\).

For each trained model, we summarize prediction stability by averaging the
interval widths over all nodes and unseen-test assignments.
% \[
% \overline{w}
% =
% \frac{1}{|\mathcal A_{\mathrm{test}}||\mathcal V|}
% \sum_{a\in\mathcal A_{\mathrm{test}}}
% \sum_{i\in\mathcal V}
% w_{a,i}.
% \]
We additionally report the corresponding mean prediction standard deviation,
% \[
% \overline{s}
% =
% \frac{1}{|\mathcal A_{\mathrm{test}}||\mathcal V|}
% \sum_{a\in\mathcal A_{\mathrm{test}}}
% \sum_{i\in\mathcal V}
% s_{a,i},
% \]
which directly measures the variation of the spreading parameter estimates across
independent feature realizations. The procedure is repeated for each
independent training run, and Table~\ref{tab:benchmark_uncertainty_5000}
reports the mean and standard deviation of the resulting run-level summaries.

The intervals are used here as empirical measures of prediction stability,
rather than as formally calibrated confidence intervals. A narrow interval
indicates that the inferred spreading parameter remains similar when the same
latent assignment is observed through independent cascade samples, whereas a
wider interval indicates greater sensitivity to the particular cascade
realization.

Table~\ref{tab:benchmark_uncertainty_5000} shows that observation noise
consistently increases this variability. Both the mean interval width and the
mean prediction standard deviation are larger in the noisy setting for every
graph, indicating that spreading parameter estimation becomes more sensitive to
the available cascade realization when the terminal observations are
corrupted.

The effect is most pronounced for the balanced tree. Its mean interval width
increases from \(0.0429\) in the noise-free setting to \(0.1154\) under noisy
observations, while the mean prediction standard deviation increases from
\(0.0346\) to \(0.0931\). This is also the graph for which the largest
deterioration in point-estimation performance is observed under noise. A
possible contributing factor is the limited structural redundancy of the
tree. Message-passing GNNs construct node representations by aggregating
information from neighboring nodes
\citep{gilmer2017neural,hamilton2017inductive}; in a tree, information between
two nodes is transmitted along a unique path, whereas graphs containing
cycles provide alternative routes through which related observations can
contribute to the representation. Consequently, corrupted observations may
have a stronger effect on the inferred susceptibilities in the tree. We
interpret this explanation cautiously, since the present experiments do not
isolate topology from the other factors affecting prediction difficulty.

The BA graphs are more stable under the same observation noise. In particular,
BA(2) and BA(3) yield the smallest noisy mean interval widths, \(0.0576\) and
\(0.0582\), respectively, while BA(4) and Karate Club lie between these graphs
and the balanced tree. The results therefore suggest that the effect of
observation noise depends not only on the level of corruption but also on the
network structure through which the observed information is aggregated.

In the simulated benchmark, the true susceptibilities are available and the
stability measures can be interpreted together with the corresponding
prediction errors. In an observed hidden-cascade system, however, the latent
susceptibilities are unknown. The same stability calculation remains
available because it depends only on repeated cascade observations and the
resulting model predictions. It therefore provides a practical indication of
whether the inferred spreading parameter vector is consistently recovered from
independent realizations of the observed spreading process.

\begin{table*}[ht!]
\centering
\small
\caption{Empirical prediction stability of SAGE-HC  corresponding to 5000 spreading parameter assignments,
10 feature realizations per assignment, and 50 cascades per seed.
Values are reported as mean $\pm$ standard deviation across 5 independent
training runs.}
\label{tab:benchmark_uncertainty_5000}
\renewcommand{\arraystretch}{1.15}
\setlength{\tabcolsep}{5pt}
\begin{tabular}{lcccc}
\toprule
\multirow{2}{*}{\textbf{Graph}} &
\multicolumn{2}{c}{\textbf{No noise}} &
\multicolumn{2}{c}{\textbf{Noisy}} \\
\cmidrule(lr){2-3} \cmidrule(lr){4-5}
& \textbf{Mean interval width} & \textbf{Mean pred. std}
& \textbf{Mean interval width} & \textbf{Mean pred. std} \\
\midrule
Balanced tree
& $0.04291 \pm 0.00015$
& $0.03462 \pm 0.00012$
& $0.11540 \pm 0.00057$
& $0.09309 \pm 0.00046$ \\

Karate Club
& $0.04886 \pm 0.00037$
& $0.03941 \pm 0.00030$
& $0.07294 \pm 0.00038$
& $0.05884 \pm 0.00030$ \\

BA(2)
& $0.03603 \pm 0.00015$
& $0.02907 \pm 0.00012$
& $0.05761 \pm 0.00031$
& $0.04648 \pm 0.00025$ \\

BA(3)
& $0.04182 \pm 0.00016$
& $0.03374 \pm 0.00013$
& $0.05819 \pm 0.00023$
& $0.04694 \pm 0.00018$ \\

BA(4)
& $0.04625 \pm 0.00045$
& $0.03731 \pm 0.00037$
& $0.06641 \pm 0.00046$
& $0.05357 \pm 0.00037$ \\
\bottomrule
\end{tabular}
\end{table*}
\subsection{Sensitivity Analysis}
\label{app:sensitivity}

For the sensitivity analysis, we consider BA(2) graphs and vary one experimental
factor at a time while keeping the remaining settings fixed. Unless stated
otherwise, we use 5,000 spreading parameter assignments, 10 feature realizations per
assignment, 50 cascades per seed, and 100 nodes. Results are reported for both
noise-free observations and the noisy setting with $p_1=p_2=0.2$ and $q=0.6$.
We report Smooth L1 error and Acc@0.1 to access the performance.

\begin{figure*}[t]
    \centering
    \includegraphics[width=0.86\textwidth]
    {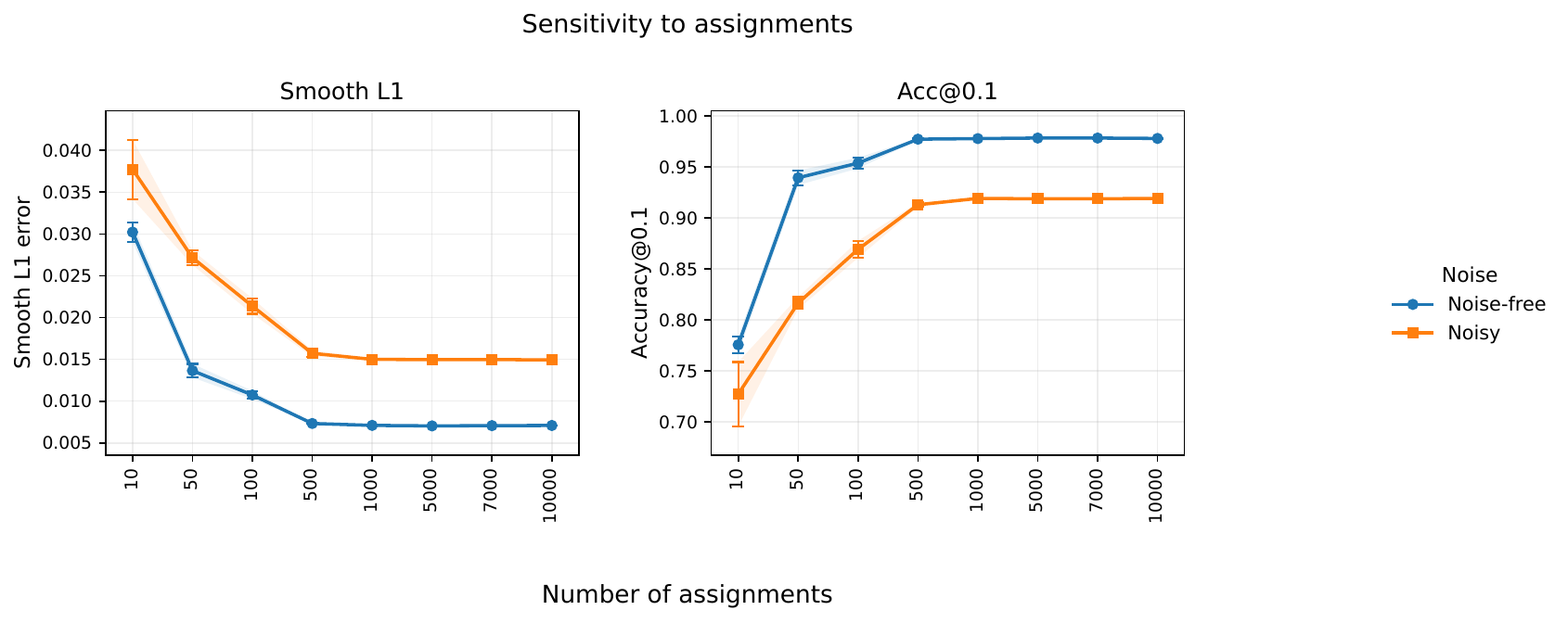}
    \caption{
    Sensitivity of SAGE-HC to the number of spreading parameter assignments on
    BA(2), with the number of feature realizations and cascades per seed held
    fixed. Increasing the number of assignments substantially improves
    estimation at small and moderate sample sizes, after which both Smooth L1
    error and Acc@0.1 approach a plateau. The noisy regime remains more
    challenging throughout. Error bars denote the standard deviation across
    repeated training runs.
    }
    \label{fig:assignment_sensitivity}
\end{figure*}

\textbf{Number of assignments:}
Figure~\ref{fig:assignment_sensitivity} examines the effect of the number of
spreading parameter assignments while keeping the graph, number of feature
realizations, and cascade budget fixed. Increasing the number of assignments
consistently reduces Smooth L1 error and improves Acc@0.1 in both observation
regimes. The largest gains occur between 10 and approximately 500 assignments,
after which both metrics stabilize. This indicates a conventional learning-curve
effect: additional assignments improve coverage of the latent spreading parameter
space, but their marginal benefit decreases once this space is sufficiently
represented. Similar saturation with increasing training-set size is commonly
observed in neural-network scaling studies~\citep{rosenfeld2019constructive}.

The effect is particularly pronounced under noisy observations. Acc@0.1
increases from approximately $0.73$ with 10 assignments to approximately $0.92$
once the number of assignments reaches $10^3$, while Smooth L1 decreases from
approximately $3.8\times10^{-2}$ to $1.5\times10^{-2}$. In the noise-free
setting, Acc@0.1 approaches $0.98$ and Smooth L1 decreases to approximately
$7\times10^{-3}$. Beyond roughly 500--1,000 assignments, further increases up
to 10,000 produce only marginal changes. 

\begin{figure*}[t]
    \centering
    \includegraphics[width=0.86\textwidth]
    {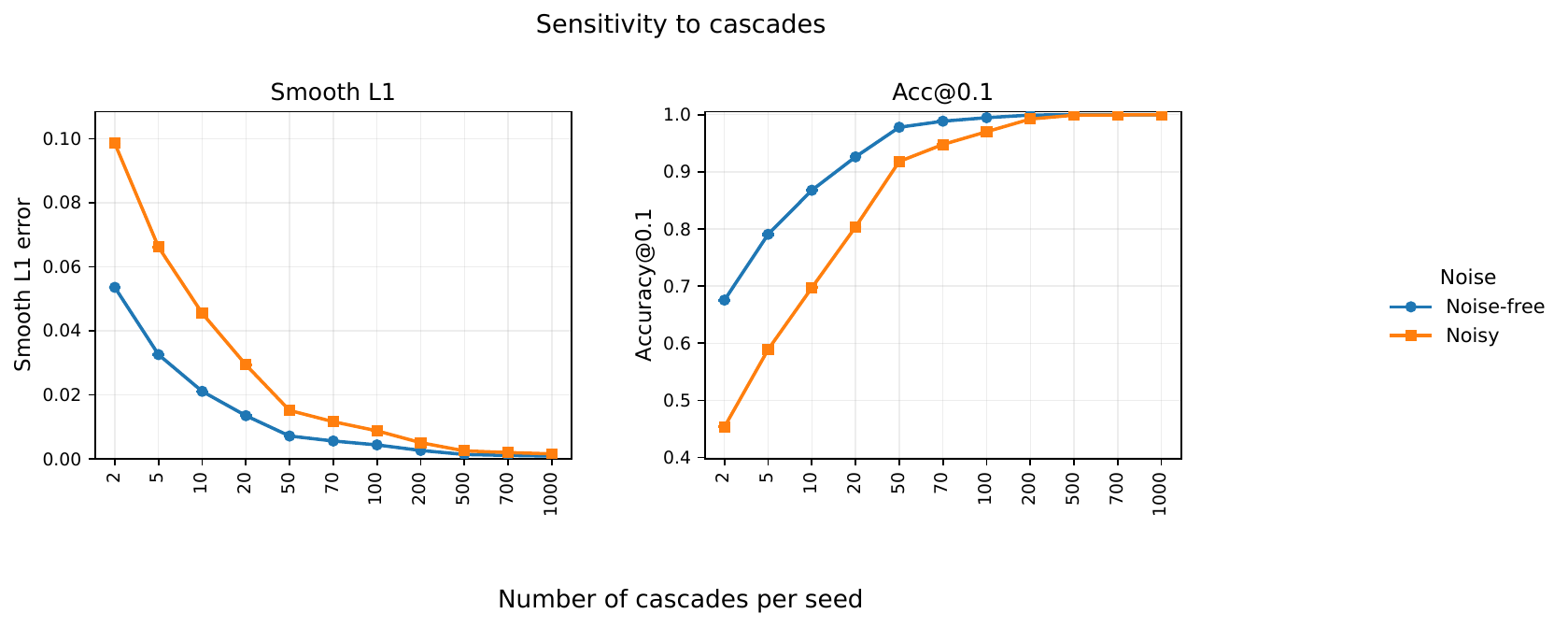}
    \caption{
    Sensitivity of SAGE-HC to the number of cascades per seed on BA(2), with
    the number of spreading parameter assignments and feature realizations held
    fixed. Increasing the cascade budget sharply reduces Smooth L1 error and
    improves Acc@0.1, with the largest gains occurring at low and moderate
    cascade counts. At sufficiently large cascade budgets, the noisy and
    noise-free regimes both approach saturation, showing that additional
    spreading evidence can substantially mitigate observation noise. Error
    bars denote the standard deviation across repeated training runs.
    }
    \label{fig:cascade_sensitivity}
\end{figure*}

\textbf{Number of cascades per seed:}
Figure~\ref{fig:cascade_sensitivity} isolates the effect of the observation
budget by varying the number of cascades generated per seed. Cascade count has a pronounced effect on estimation
performance. Smooth L1 decreases monotonically as the number of cascades
increases from 2 to 1,000, while Acc@0.1 rapidly approaches one. The largest
improvement occurs in the low-to-moderate cascade regime, showing that a small
number of cascades provides insufficiently stable evidence for accurate
spreading parameter estimation.

The effect is substantially stronger under noisy observations. With only two
cascades per seed, Acc@0.1 is approximately $0.45$ in the noisy setting,
compared with approximately $0.68$ in the noise-free setting. As the cascade
budget increases, this gap progressively narrows. At approximately 200 cascades
per seed, both regimes already achieve very high Acc@0.1, and at 500--1,000
cascades their accuracies are effectively saturated near one. Smooth L1 exhibits
the same convergence, with both curves approaching very small errors at the
largest cascade budgets.

These results show that observation noise is strongly information-limited:
additional spreading realizations can compensate for the reduced reliability of
individual noisy cascades. This behavior is consistent with the broader
sample-complexity literature on cascade-based inverse problems, where increasing
the number of independent spreading observations improves recoverability and
parameter identification~\citep{daneshmand2014estimating,
pouget2015inferring}. Although these works address network-structure
recovery rather than node-level spreading parameter regression, the underlying
information-budget principle is closely related. The flattening of both curves
at large cascade counts further demonstrates diminishing returns once the
cascade-derived statistics are estimated with sufficient precision.

\begin{figure*}[t]
    \centering
    \includegraphics[width=0.86\textwidth]
    {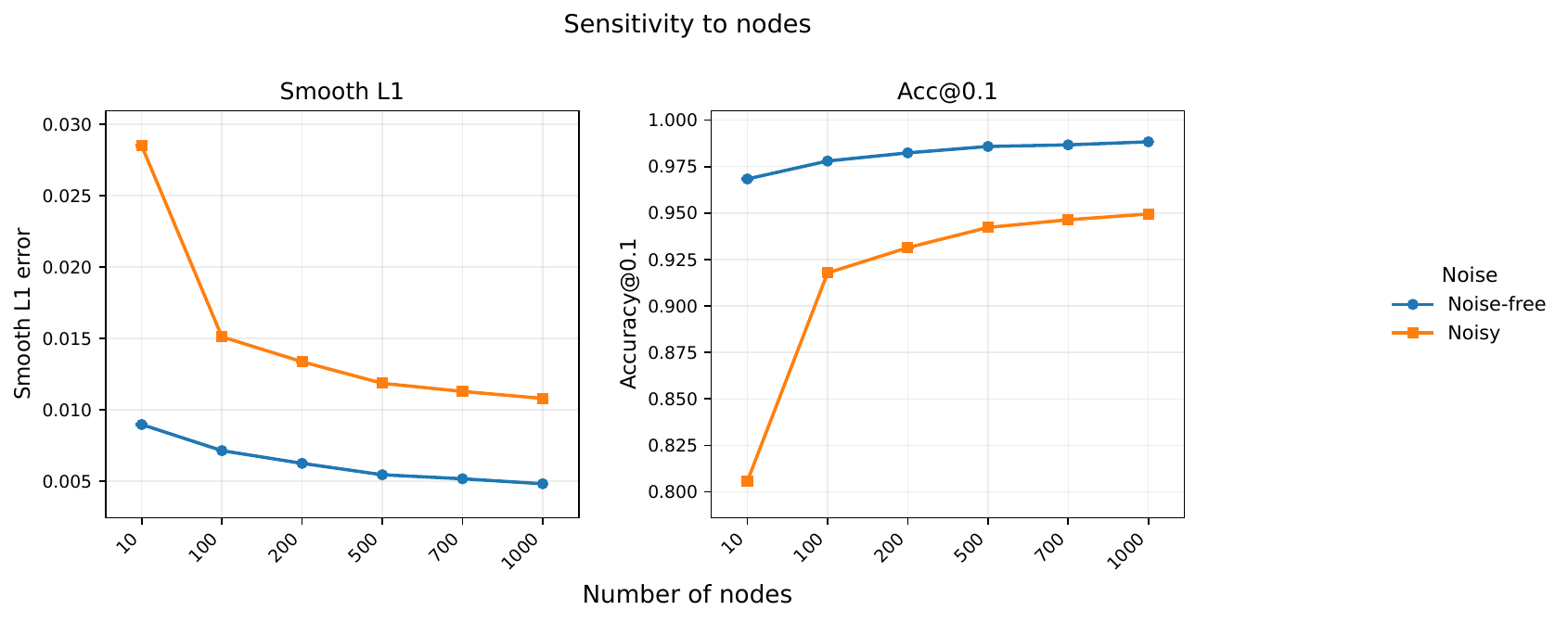}
    \caption{
    Sensitivity of SAGE-HC to graph size on BA(2), with the number of
    spreading parameter assignments, feature realizations per assignment, and
    cascades per seed held fixed. Performance improves with increasing graph
    size, particularly under noisy observations, and exhibits diminishing
    returns for larger graphs. Since all nodes are used as seeds, the experiment
    measures end-to-end scaling under a fixed per-seed cascade budget rather
    than graph size under a fixed total observation budget. Error bars denote
    the standard deviation across repeated training runs where multiple runs
    are available.
    }
    \label{fig:node_sensitivity}
\end{figure*}

\textbf{Number of nodes:}
Figure~\ref{fig:node_sensitivity} evaluates the behavior of SAGE-HC as the
BA(2) graph size increases from 10 to 1,000 nodes. Performance does not deteriorate as the node-level prediction problem becomes
larger. Instead, Smooth L1 decreases monotonically and Acc@0.1 increases over
the tested graph sizes. The improvement is modest in the noise-free regime,
where performance is already high for the smallest graph, but is considerably
larger under noisy observations.

The strongest improvement occurs between 10 and 100 nodes. In the noisy
setting, Acc@0.1 increases from approximately $0.81$ to $0.92$, while Smooth L1
decreases from approximately $2.9\times10^{-2}$ to $1.5\times10^{-2}$.
Performance continues to improve as the graph grows, but at a progressively
smaller rate, reaching an Acc@0.1 of approximately $0.95$ at 1,000 nodes. In
the noise-free setting, Acc@0.1 increases from approximately $0.97$ to $0.99$,
while Smooth L1 falls below $5\times10^{-3}$ at the largest graph size.
Consequently, increasing the prediction dimensionality from 10 to 1,000 nodes
does not lead to a loss of estimation accuracy within the investigated regime.

The node-sensitivity experiment should, however, be interpreted as an
end-to-end scaling experiment under a fixed \emph{per-seed} observation budget,
rather than as an isolated effect of graph size. Since all nodes are used as
seeds, increasing the number of nodes also increases the total number of
seed-conditioned cascade observations contributing to each feature realization.
The observed improvement can therefore reflect both the richer relational
structure of larger graphs and the greater aggregate amount of spreading
evidence. Importantly, neither effect results in degradation as the number of
node-level parameters increases. The progressively smaller gains beyond
approximately 200--500 nodes indicate that the benefit of additional graph
scale eventually exhibits diminishing returns.

\textbf{Effect of training-time feature realizations:}
To examine the role of repeated stochastic feature realizations, we vary the
number of realizations available during training,
\(F_{\mathrm{train}}\in\{2,5,10,20,50,70,100\}\), while fixing the strict
unseen-test budget at \(F_{\mathrm{test}}=2\) for every model. This controlled
design ensures that all models are evaluated using the same amount of
test-time information, so differences in performance can be attributed to the
number of feature realizations used during training rather than to an
increasing observation budget at inference.

\begin{figure*}[h]
    \centering
    \includegraphics[width=\textwidth]
    {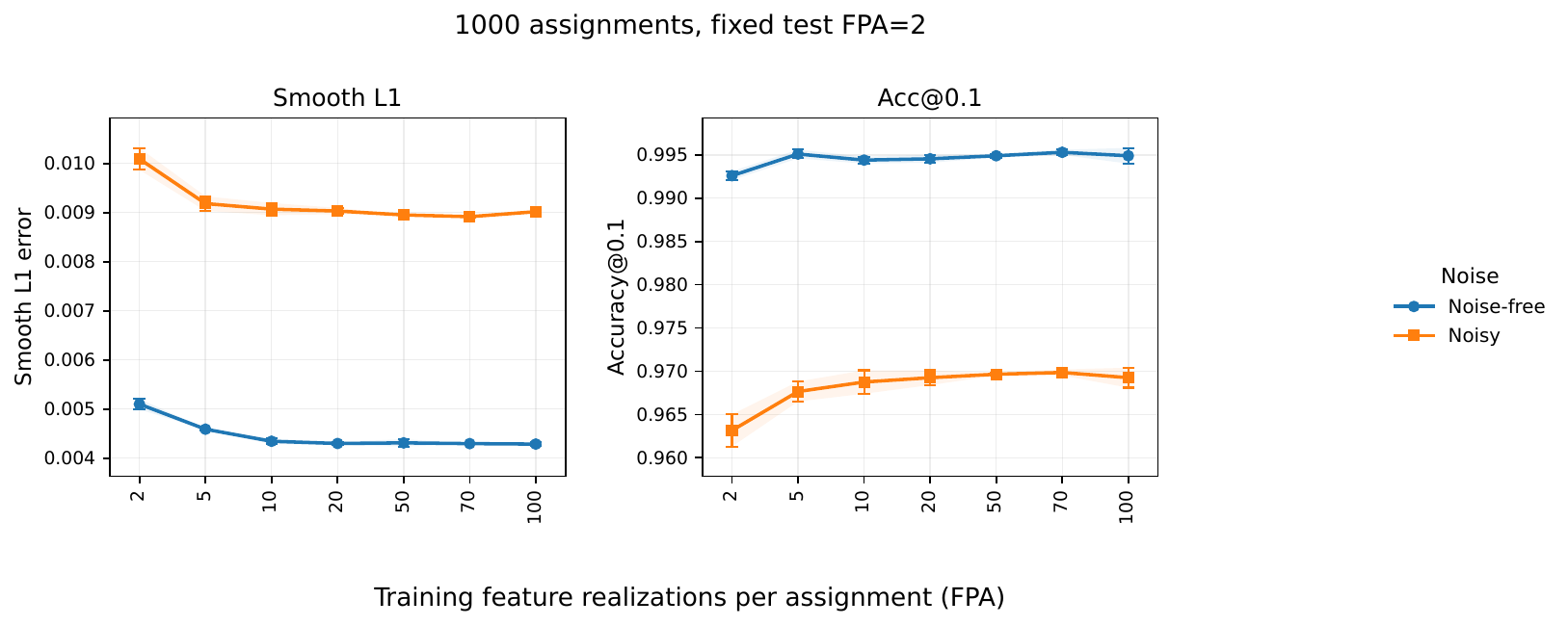}
    \caption{
    Effect of training-time feature replication on BA(2) with 1,000
spreading parameter assignments. The number of feature realizations used during
training is varied, while the strict unseen-test budget is fixed at
\(F_{\mathrm{test}}=2\) for every model. Fixing the test budget isolates the
effect of training-time replication from changes in the amount of
information available at inference. Performance improves primarily for small
to moderate training FPA, particularly under hidden/noisy observations, and
shows diminishing returns thereafter. Error bars denote variability across
independent training runs.
    }
    \label{fig:fpa_fixed_test_1000}
\end{figure*}

Figure~\ref{fig:fpa_fixed_test_1000} reports the results for 1,000
spreading parameter assignments. Increasing the training FPA from 2 to a modest
number of repeated realizations improves predictive performance in both
observation regimes. In the noise-free setting, Smooth L1 decreases from
approximately \(0.0051\) at \(F_{\mathrm{train}}=2\) to about \(0.0043\) by
\(F_{\mathrm{train}}=20\). Under hidden/noisy observations, the improvement is
more pronounced: Smooth L1 decreases from approximately \(0.0101\) to around
\(0.0091\), while Acc@0.1 increases from approximately \(0.96\) to about \(0.970\).

Beyond approximately 10--20 training realizations, both performance measures
largely stabilize. Increasing \(F_{\mathrm{train}}\) further to 50, 70, or
100 yields only marginal additional gains. This behavior indicates that a
small number of repeated realizations is sufficient to expose the model to
the stochastic variability associated with a fixed spreading parameter assignment,
whereas additional realizations become increasingly redundant once this
variability has been adequately represented. The effect is more visible in
the hidden/noisy regime, where repeated realizations are particularly useful
for learning a mapping that is robust to observation corruption.

\textbf{Generalization across test-time observation budgets:} We further evaluate whether an observation-rich model remains effective when
less cascade evidence is available at inference. We use 5,000 spreading parameter
assignments on BA(2), of which 4,500 are allocated to training and validation
and the remaining 500 are held out for testing. Each assignment contains 10
feature realizations, and SAGE-HC is trained once with
\(C_{\mathrm{train}}=1000\) cascades per seed. After checkpoint
selection, the model is frozen and evaluated at
\(C_{\mathrm{test}}\in\{5,10,50,100,500,700,1000\}\).

\begin{figure*}[t]
    \centering
    \includegraphics[width=\textwidth]
    {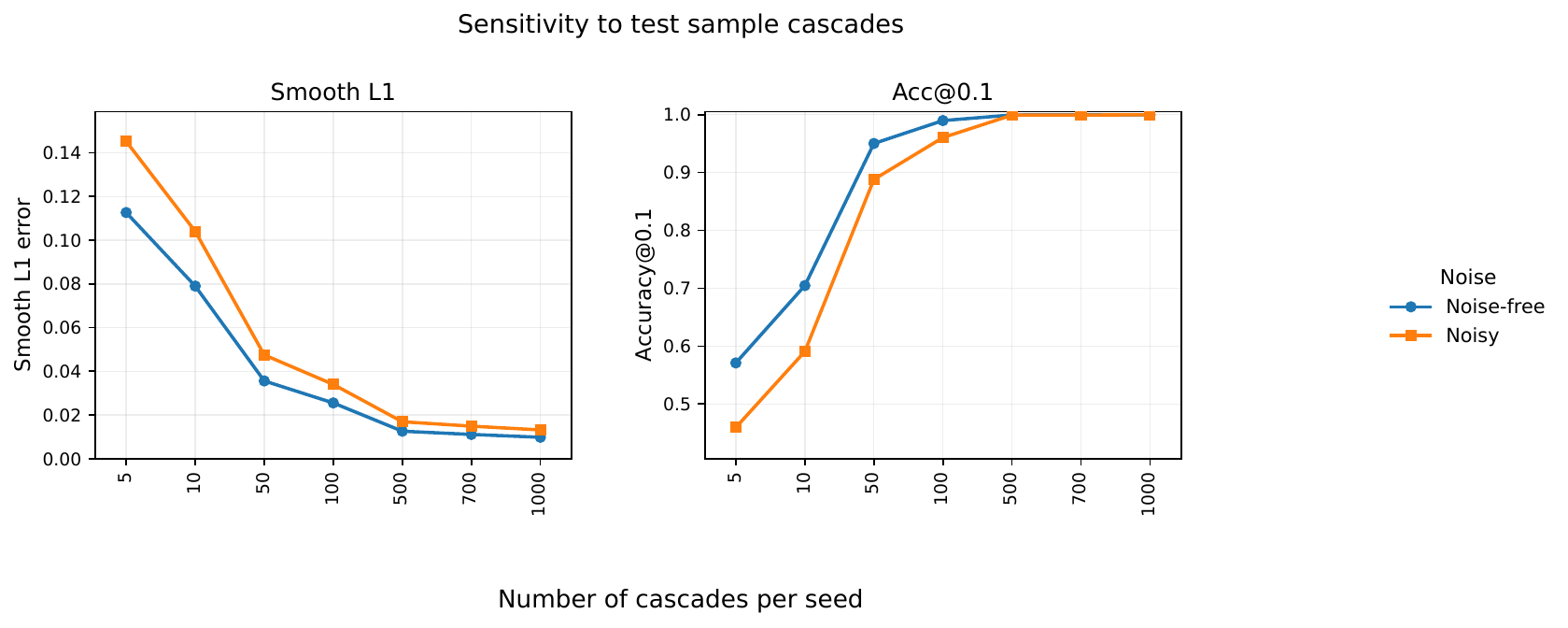}
    \caption{
    Sensitivity to the number of cascades available at test time for a frozen
    SAGE-HC model on BA(2). The model is trained once with \(C_{\mathrm{train}}=1000\) cascades per seed. The same held-out test
    assignments are then evaluated using different test-time cascade budgets.
    The left panel reports prediction error and the right panel reports
    Acc@0.1. Performance improves rapidly as additional cascade evidence becomes
    available, particularly under hidden/noisy observations, and approaches a
    plateau at larger test-time budgets.
    }
    \label{fig:cross_cascade_generalization}
\end{figure*}

Figure~\ref{fig:cross_cascade_generalization} examines the robustness of a
frozen SAGE-HC model to changes in the amount of cascade evidence available
at test time. At inference time, the same
held-out assignments are evaluated with progressively smaller observation
budgets, ranging from 5 to 1,000 cascades per seed. Thus, unlike the preceding
cascade-sensitivity experiment, the model is not retrained for each cascade
budget; only the test-time observation budget is varied.

Performance depends strongly and monotonically on the amount of test-time
cascade evidence. In the noise-free setting, the prediction error decreases
rapidly as \(C_{\mathrm{test}}\) increases, while Acc@0.1 rises from
approximately \(0.57\) with only five cascades per seed to about \(0.95\) at
50 cascades and nearly \(0.99\) at 100 cascades. Further increases yield
progressively smaller improvements, with accuracy approaching unity at the
largest observation budgets.

The hidden/noisy setting exhibits the same qualitative behavior but is more
strongly affected when only a small number of cascades is available.
Acc@0.1 is approximately \(0.46\) at five cascades per seed and \(0.59\) at
10 cascades, compared with approximately \(0.57\) and \(0.70\),
respectively, in the noise-free setting. The gap narrows substantially as
additional observations become available: noisy Acc@0.1 reaches about
\(0.89\) at 50 cascades and \(0.96\) at 100 cascades, and the two regimes are
nearly indistinguishable at the largest cascade budgets.

These results show that the learned inverse mapping transfers across
substantially different test-time observation budgets without retraining.
At very small cascade counts, estimation is information-limited because the
cascade-derived representation is based on insufficient evidence, an effect
that is amplified by hidden and corrupted observations. Once a moderate
number of cascades is available, however, performance improves rapidly and
then exhibits diminishing returns. The experiment therefore separates
\emph{training-time data requirements} from \emph{test-time evidence
requirements} and shows that a model trained under an observation-rich
setting remains effective when substantially less cascade information is
available at inference time.

\begin{figure*}[t]
\centering
\includegraphics[width=1\textwidth]{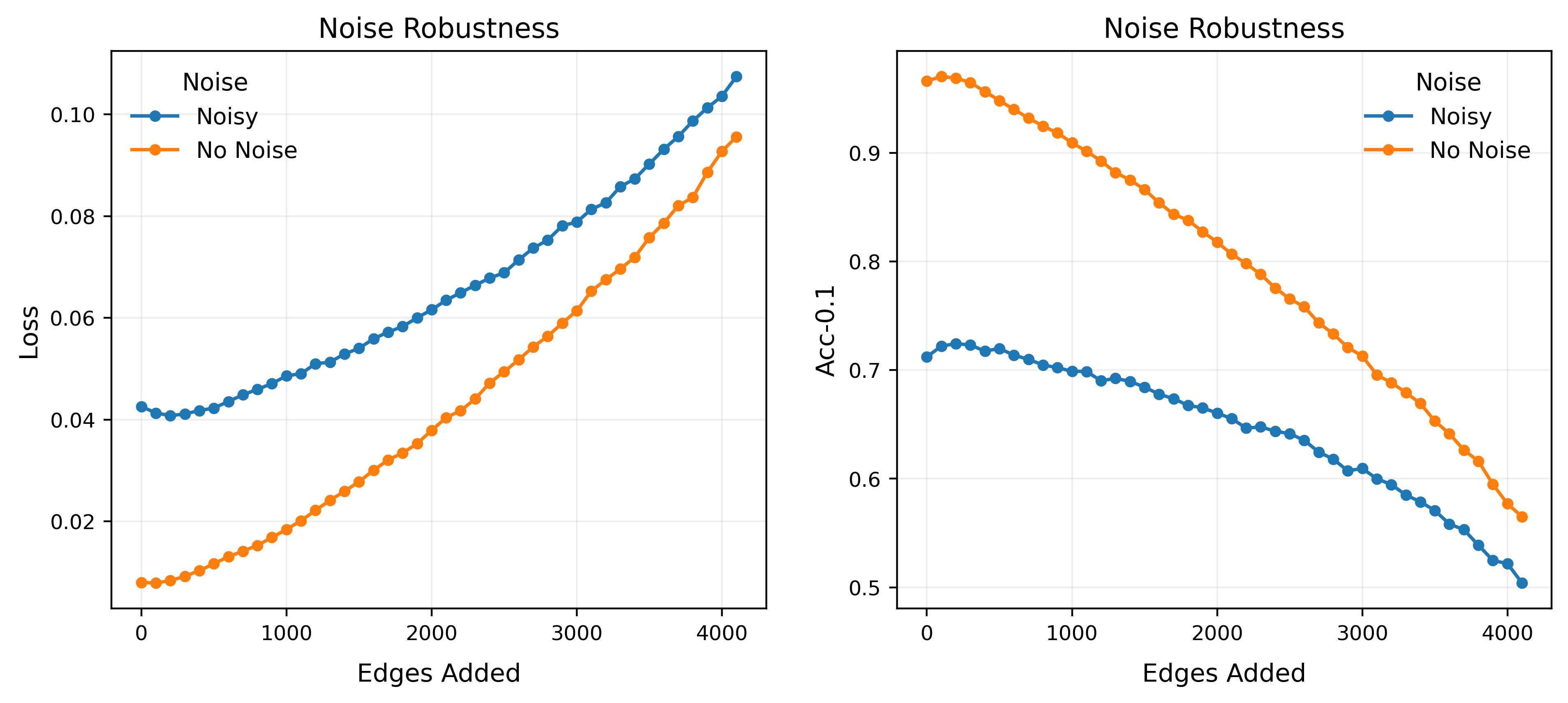}
\caption{Comparison of accuracies at the $0.1$ threshold of the noise-free and noisy training settings for a tree graph with progressively added edges to increase its loopiness}
\label{fig:progressive_tree}
\end{figure*}

\textbf{Noise Robustness:}
We analyze how the loopiness of the graph influences the  noise robustness of the model. To implement this, we start with a $100$-node tree graph and gradually add edges to increase loopiness. Figure \ref{fig:progressive_tree} compares the accuracy at the $0.1$ threshold of the models under the noise-free and noisy scenarios.
The accuracy difference starts at $22.82$\% for the original tree graph and gradually decreases with more edges added. For a graph with $4100$ edges added, the accuracy difference is just $6.11$\% but the total accuracy is also lower, with the noise-free accuracy being $56.50$\%.
Both no-noise and noisy accuracies improve for the first 100-200 edges added and then decline steadily. 
This experiment indicates that the graph's loopiness directly influences the robustness to noise of the model.

\end{document}